\documentclass[12pt]{article}

\usepackage[T1]{fontenc}
\usepackage[title]{appendix}
\usepackage{amsmath}
\usepackage{amssymb}
\usepackage{slashed}
\usepackage{tikz}
\usepackage[numbers,sort&compress]{natbib}
\usepackage{hyperref}
\usepackage{cleveref}
\crefname{equation}{Eq.}{Eqs.}
\crefname{figure}{Fig.}{Figs.}
\crefname{table}{Table}{Tables}
\crefname{section}{Section}{Sections}

\usepackage[letterpaper,margin=1in,bottom=1in]{geometry}
\usepackage{float} % allows forcing plot and table locations with [H]
\usepackage{parskip} % skip spaces between paragraphs
\usepackage{tabulary} % table environment with auto word wrapping
\usepackage{color} % color for tracking edits
\usepackage{soul} % strikeout for tracking edits
\usepackage{subfig}
\usepackage{graphicx}
\usepackage[section]{placeins} % keep figures inside their sections
\usepackage{multirow}

\def\lhc2{LHC~Run~II}

\newcommand{\code}[1]{\texttt{#1}}

\numberwithin{equation}{section}
\def\beq{\begin{equation}}
\def\be{\begin{equation}}
\def\beqn{\begin{eqnarray}}
\def\ee{\end{equation}}
\def\eeq{\end{equation}}
\def\eeqn{\end{eqnarray}}

\def\r2{\sqrt 2}

\begin{document}

\author{Amin Aboubrahim$^a$\footnote{Email: aabouibr@uni-muenster.de}, ~Tarek Ibrahim$^{b}$\footnote{Email: tibrahim@zewailcity.edu.eg}, ~Michael Klasen$^a$\footnote{Email: michael.klasen@uni-muenster.de}, ~and Pran Nath$^c$\footnote{Email: p.nath@northeastern.edu}\\~\\
$^{a}$\textit{\normalsize Institut f\"ur Theoretische Physik, Westf\"alische Wilhelms-Universit\"at M\"unster,} \\ 
\textit{\normalsize Wilhelm-Klemm-Stra{\ss}e 9, 48149 M\"unster, Germany}\\
$^b$\textit{\normalsize University of Science and Technology, Zewail City of Science and Technology,} \\
\textit{\normalsize 6th of October City, Giza 12588, Egypt} \\
$^{c}$\textit{\normalsize Department of Physics, Northeastern University, Boston, MA 02115-5000, USA}} 
\title{\vspace{-2.0cm}\begin{flushright}
{\small MS-TP-20-45}
\end{flushright}
\vspace{1cm}
A decaying neutralino as dark matter and its gamma ray spectrum}
\date{}
\maketitle

\begin{abstract}

It is shown that a decaying neutralino in a supergravity unified framework is a viable candidate for dark
matter. Such a situation arises in the presence of a hidden sector with ultraweak couplings to the visible sector where the neutralino can decay into the hidden sector's lightest supersymmetric particle (LSP) with a  lifetime larger than the lifetime of the universe. We present a concrete model  where the MSSM/SUGRA 
is extended to include a hidden sector comprised of $U(1)_{X_1} \times U(1)_{X_2}$ gauge sector
and the LSP of the hidden sector is a neutralino which is lighter than the LSP neutralino of the visible sector.
We compute  the loop suppressed radiative decay of the visible sector neutralino into the neutralino of the hidden sector  and show that the decay can occur with a lifetime larger than the age of the universe. 
  The decaying neutralino can be probed by indirect detection experiments,
specifically by its signature decay into the hidden sector neutralino and an energetic gamma ray photon.
Such a gamma ray can be searched for with improved sensitivity
 at  Fermi-LAT and by future experiments such as the Square Kilometer Array (SKA) and the 
Cherenkov Telescope Array (CTA).  
 We present several benchmarks which have a natural suppression of the hadronic channels from dark matter annihilation and decays and consistent with measurements of the antiproton background.
\end{abstract}

\section{Introduction}\label{sec:intro}

In supergravity (SUGRA) unified models and in string based models with conserved $R$-parity, the lightest supersymmetric particle
(LSP) is absolutely stable and if neutral is a candidate for dark matter (DM).   In extended 
supergravity and string models with  hidden sectors, the LSP may lie in one of the hidden sectors
and if there is a coupling between the visible and the hidden sectors, then the LSP of the visible 
sector will be unstable and decay into the hidden sector LSP. 
Although the visible sector LSP is unstable, it could still be the dominant component of dark matter
today if its lifetime is much larger than the age of the universe. Such a situation can 
come about if the coupling of the visible sector with the hidden sector is ultraweak. 
Such a dark matter would be detectable by its energetic gamma ray signature following its decay or annihilation into photons. We are interested in examining gamma ray emissions due to DM decay into a photon and a neutral particle which produces a 
  sharp monoenergetic spectral line which could be observed over the isotropic diffuse background. Further, here one can easily trace the gamma ray signal back to its source since photons preserve spectral and spatial information unlike antimatter which suffer from energy losses and diffusion as they propagate in the universe. Also, unlike annihilation, DM decay signals do not depend on the halo substructures and hence are prone to less uncertainties in this regard. In our analysis we use the most recent astrophysical constraints on the gamma ray flux, the positron excess and the antiproton flux to isolate a parameter space of models which still evade those constraints while can be probed by future experiments looking for excess gamma ray flux in the sky.

 In this work we present a model
where the LSP of the visible sector is a neutralino  and has a radiative decay into the LSP of the 
hidden sector producing a monochromatic gamma ray signal. 
 We consider this phenomenon 
in the framework of a gauged $U(1)_{X_1}\times U(1)_{X_2}$ extended supergravity unified model. 
 The $U(1)_{X_1}$ sector has kinetic coupling with the visible sector (characterized by $\delta_1$)
 while the $U(1)_{X_2}$ sector 
 has no kinetic coupling with the visible sector and has only kinetic coupling with the $U(1)_{X_1}$ 
 sector (characterized by $\delta_2$).   After a transformation to the canonical kinetic energy frame, one finds that the visible sector
 develops a tiny coupling to the hidden sector $U(1)_{X_2}$ which is proportional to the product 
 of the kinetic couplings, i.e., $\delta_1\delta_2$.  Furthermore, the $U(1)_{X_1}$ and $U(1)_{X_2}$  
 gauge multiplets gain masses via the Stueckelberg mechanism. We assume that the lightest neutralino
 ($\tilde\xi_1^0$)  resides in the hidden sector $X_2$ and has ultraweak interactions with the neutralino 
of the visible sector. Assuming that the LSP of the visible sector is the lightest neutralino
 in the visible sector ($\tilde\chi_1^0$), and assuming $R$-parity conservation, 
 one will have a loop induced decay $\tilde\chi_1^0\to \tilde\xi_1^0\gamma$.
We present 
a set of benchmarks within this extended SUGRA grand unified model where the relic density of
$\tilde\chi_1^0$ is achieved through the normal thermal freeze-out mechanism. On the other hand, the neutralino $\tilde \xi_1^0$ residing in the hidden sector $X_2$ has interactions too feeble to be produced in the early universe to any
discernible amount and the only production mechanism for it is via the decay of 
$\tilde\chi_1^0$. Since the lifetime of $\tilde\chi_1^0$ is much larger than the lifetime of the universe,
the contribution of $\tilde\xi_1^0$ to the relic density is negligible compared to
that of $\tilde\chi_1^0$. Thus, although unstable, the visible sector neutralino 
$\tilde\chi_1^0$
is the dominant component of dark matter in this model.

The outline of the rest of the paper is as follows: In section~\ref{sec:model}, we describe the 
$U(1)_{X_1}\times U(1)_{X_2}$ extended SUGRA model and exhibit the gauge kinetic mixing between the 
visible sector and the hidden sectors and also describe the Stueckelberg mass growth
for the gauge fields and the gauginos in the hidden sectors.  In section~\ref{sec:loopwidth}, we give details
of the loop analysis of the radiative decay of the visible sector neutralino which involves computation of supersymmetric  diagrams with $W$-charginos, charged Higgs-charginos, and fermions-sfermions in the loops. Specifically we compute the electric and magnetic transition dipole moments that enter in the decay width arising from the diagrams of  Fig.~\ref{fig1}. Details of the 
numerical analysis are given in section~\ref{sec:numerics} and the conclusion in section~\ref{sec:conc}. In Appendix~\ref{sec:loop}, we list the interactions
that enter in the analysis of radiative decay of the neutralino.
In Appendix~\ref{sec:id}, we give an analysis of the gamma ray and antiproton flux calculations.

%The antiproton flux arising from the annihilation and three body decay of the neutralino is analyzed in section~\ref{sec:antipflux}. Here it is shown that the contribution to the antiproton
%flux generated by the model points is negligible consistent with the background 
%antiproton flux from astrophysical sources. Calculations of the isotropic gamma ray background are presented in section~\ref{sec:igrb}.

\section{The model}\label{sec:model}

 As mentioned in the introduction, we consider a SUGRA grand unified model 
 extended  with extra $U(1)_{X_1}$ and $U(1)_{X_2}$ hidden gauge sectors. One of the gauge superfields of this
 model $\hat X_1$   mixes with the hypercharge gauge superfield $\hat B$ where their components
 in the Wess-Zumino gauge are given by $\hat X_1 =(X_1^\mu, \lambda_{X_1}, D_{X_1})$ and
 $\hat B= (B^\mu, \lambda_B, D_B)$ while the other gauge superfield $\hat X_2$ with components $\hat X_2 =(X_2^\mu, \lambda_{X_2}, D_{X_2})$  mixes directly only with $\hat X_1$. We assume a kinetic mixing~\cite{Holdom:1985ag}
 between $U(1)_{X_1}$ and $U(1)_Y$  and between $U(1)_{X_1}$ and $U(1)_{X_2}$ so the  kinetic part of the Lagrangian is given by
\begin{align}
\mathcal{L}_{\rm gk}=&-\frac{1}{4}(B_{\mu\nu}B^{\mu\nu}+X_{1\mu\nu}X_1^{\mu\nu}+X_{2\mu\nu}X_2^{\mu\nu})-i\lambda_B\sigma^{\mu}\partial_{\mu}\bar{\lambda}_B-i\lambda_{X_1}\sigma^{\mu}\partial_{\mu}\bar{\lambda}_{X_1}-i\lambda_{X_2}\sigma^{\mu}\partial_{\mu}\bar{\lambda}_{X_2} \nonumber \\
&+\frac{1}{2}(D^2_B+D^2_{X_1}+D^2_{X_2})-\frac{\delta_1}{2}B^{\mu\nu}X_{1\mu\nu}-\frac{\delta_2}{2}X_1^{\mu\nu}X_{2\mu\nu}+\delta_1 D_B D_{X_1}+\delta_2 D_{X_1} D_{X_2} \nonumber \\
&-i\delta_1(\lambda_{X_1}\sigma^{\mu}\partial_{\mu}\bar{\lambda}_B+\lambda_{B}\sigma^{\mu}\partial_{\mu}\bar{\lambda}_{X_1})-i\delta_2(\lambda_{X_1}\sigma^{\mu}\partial_{\mu}\bar{\lambda}_{X_2}+\lambda_{X_2}\sigma^{\mu}\partial_{\mu}\bar{\lambda}_{X_1}).
\label{kinetic-2}
\end{align}
In addition, we assume a Stueckelberg Lagrangian which induces a mass growth for the hidden
sector~\cite{Kors:2004dx,Cheung:2007ut,Feldman:2007wj} so that
\begin{equation}
\mathcal{L}_{\rm St}=\int d\theta^2 d\bar{\theta}^2\left[(M_1 X_1+S_1+\bar{S_1})^2+(M_2 X_2+S_2+\bar{S_2})^2\right],
\label{lag}
\end{equation}
where $S_1$, $\bar S_1$ and $S_2$, $\bar S_2$ are chiral superfields and their presence guarantees gauge invariance
of Eq.~(\ref{lag}) under $U(1)_Y$, $U(1)_{X_1}$ and $U(1)_{X_2}$ gauge transformations, where 
$M_1 (M_2)$ give mass of the hidden sector field $X_1 (X_2)$.
The components of the chiral fields $\hat S_i$ ($i=1,2$)  are $\hat S_i=(\rho_i+ ia_i, \chi_i, F_i)$, where 
$\rho_i+ i a_i$ are the chiral scalars in $\hat S_i$, $\chi_i$ are the chiral fermions and $F_i$ the auxiliary fields
and a similar component form holds for the superfields $\hat{\bar S}_i$. 
Further, in component notation, $\mathcal{L}_{\rm St}$ is given by
\begin{align}
\mathcal{L}_{\rm St} = &-\frac{1}{2}(M_1 X_{1\mu}+\partial_{\mu}a_1)^2-\frac{1}{2}(M_2 X_{2\mu}+\partial_{\mu}a_2)^2-\frac{1}{2}\left[(\partial_{\mu}\rho_1)^2+(\partial_{\mu}\rho_2)^2\right] \nonumber \\
&-i\chi_1\sigma^{\mu}\partial_{\mu}\bar{\chi_1}-i\chi_2\sigma^{\mu}\partial_{\mu}\bar{\chi_2}+2(|F_1|^2+|F_2|^2)+M_1\rho_1 D_{X_1}+M_2\rho_2 D_{X_2} \nonumber \\
&+M_1(\bar{\chi_1}\bar{\lambda}_{X_1}+\chi_1\lambda_{X_1})+M_2(\bar{\chi_2}\bar{\lambda}_{X_2}+\chi_2\lambda_{X_2}).
\end{align}
In the unitary gauge, the axion fields $a_1$ and $a_2$ are absorbed to generate mass for the $U(1)_{X_1}$ and $U(1)_{X_2}$ gauge bosons.
It is convenient from this point on to introduce Majorana spinors $\psi_S$, $\Lambda_{X_1}$, $\Lambda_{X_2}$ and $\lambda_Y$ so that
 \begin{equation}
  \psi_{S_1} =
  \begin{pmatrix}
    \chi_{1\alpha}  \\
    \bar{\chi}_1^{\dot{\alpha}}
  \end{pmatrix},\quad
  \psi_{S_2} =
  \begin{pmatrix}
    \chi_{2\alpha}  \\
    \bar{\chi}_2^{\dot{\alpha}}
  \end{pmatrix},\quad
  \Lambda_{X_1}=
  \begin{pmatrix}
    \lambda_{X_1\alpha}  \\
    \bar{\lambda}^{\dot{\alpha}}_{X_1}
  \end{pmatrix},\quad
  \Lambda_{X_2}=
  \begin{pmatrix}
    \lambda_{X_2\alpha}  \\
    \bar{\lambda}^{\dot{\alpha}}_{X_2}
  \end{pmatrix},\quad
  \lambda_Y=
  \begin{pmatrix}
    \lambda_{B\alpha}  \\
    \bar{\lambda}^{\dot{\alpha}}_{B}
  \end{pmatrix}.
  \label{spinors}
\end{equation}
In addition, we include soft gaugino mass terms  so that
\begin{equation}
-\Delta\mathcal{L}_{\rm soft} \
=\frac{1}{2}m_{X_1}\bar{\Lambda}_{X_1}\Lambda_{X_1}+\frac{1}{2}m_{X_2}\bar{\Lambda}_{X_2}\Lambda_{X_2}\,.
\end{equation}
Further, we make a transformation to put the kinetic energy of the $U(1)$ fields in a canonical 
form so that the kinetic energy of the fields is diagonal and normalized. The transformation
that accomplishes this is 
%before proper implementation of the model one must carry out a diagonalization of the
%kinetic energy sector which can be done using the transformation
%Now one may check that the transformation which diagonalizes the kinetic energy Lagrangian and keeps the kinetic energy appropriately normalized is 
\begin{eqnarray}
\left(\begin{matrix} B^{\mu} \cr 
X_1^{\mu} \cr
X_2^{\mu} 
\end{matrix}\right) = \left(\begin{matrix} 1 & -s_1 & s_1 s_2 \cr 
0 & c_1 & -c_1 s_2 \cr
0 & 0 & c_2   
\end{matrix}\right)\left(\begin{matrix} B'^{\mu} \cr 
X_1'^{\mu} \cr
X_2'^{\mu}
\end{matrix}\right), 
\label{rotation}
\end{eqnarray}
where the fields
\begin{align}
c_{1}&=\frac{1}{\sqrt{1-\delta^2_1}}, ~~~~~~~~~ s_{1}=\frac{\delta_1}{\sqrt{1-\delta^2_1}}, \nonumber \\
c_{2}&=\frac{\sqrt{1-\delta^2_1}}{\sqrt{1-\delta^2_1-\delta^2_2}}, ~~~ s_{2}=\frac{\delta_2}{\sqrt{1-\delta^2_1-\delta^2_2}}.
\end{align}
Note that $c^2_{1}-s^2_{1}=1=c^2_{2}-s^2_{2}$.
%More generally we have both kinetic mixing and the Stueckelberg mass mixing~\cite{Feldman:2007wj}.
 For the MSSM/SUGRA grand unified model~\cite{mSUGRA}
 we will assume the soft sector to have non-universalities
 and characterize this sector with parameters~\cite{nonuni-gaugino,Herrmann:2009mp}
 $m_0, ~A_0, ~m_1, ~m_2, ~m_3, ~\tan\beta,
 ~\text{sgn} (\mu)$. Here
 $m_0$ is the universal scalar mass, $A_0$ is the universal trilinear coupling, $m_1,  ~m_2,  ~m_3$ are the masses of the $U(1)_Y$, $SU(2)_L$, and $SU(3)_C$ gauginos, $\tan\beta=v_u/v_d$ is the ratio of the Higgs vacuum expectation values and $\text{sgn}(\mu)$ is the sign of the Higgs mixing parameter which is chosen to be positive.
We display now the mass matrix of the neutralino  sector for SUGRA  and 
for the hidden sector in the basis  $(\lambda_Y,\lambda_3,\tilde h_1, \tilde h_2,\psi_{S_1},\Lambda_{X_1},\psi_{S_2},\Lambda_{X_2})$.
Here
 $\lambda_Y, \lambda_3, \tilde h_1, \tilde h_2$ are the gaugino and higgsino fields of the MSSM sector,
and  $\psi_{S_1}, \Lambda_{X_1}$ and $\psi_{S_2}, \Lambda_{X_2}$ are the higgsino-gaugino fields for the hidden sectors. In this basis, the $8\times 8$ MSSM/SUGRA and hidden sectors neutralino mass matrix is given by
\begin{eqnarray}
\mathcal{M}_{\chi\xi}=
\left(\begin{matrix} 
 \mathcal{M}^{\chi}_{4\times 4} & C \cr
 \overline{C} & \mathcal{M}^{\rm hid}_{4\times 4}
\end{matrix}\right),
\label{nmatrix1}
\end{eqnarray}
where $ \mathcal{M}^{\chi}_{4\times 4}$ is the standard $4\times 4$ MSSM  neutralino mass matrix given by
\begin{eqnarray}
\mathcal{M}^{\chi}_{4\times 4}=
\left(\begin{matrix} 
 m_1 & 0 & -c_{\beta}s_W M_Z & s_{\beta}s_W M_Z\cr
 0 & m_2 &
  c_{\beta}c_W M_Z & -s_{\beta}c_W M_Z \cr
-c_{\beta}s_W M_Z & c_{\beta}c_W M_Z & 0 & -\mu \cr
s_{\beta}s_W M_Z & -s_{\beta}c_W M_Z & -\mu & 0  
\end{matrix}\right),
\label{nmatrix2}
\end{eqnarray}
where
$s_{\beta}\equiv\sin\beta$, $c_{\beta}\equiv\cos\beta$, $s_W\equiv\sin\theta_W$, $c_W\equiv\cos\theta_W$ with $M_Z$ being the $Z$ boson mass and $\theta_W$ the weak mixing angle. In Eq.~(\ref{nmatrix1}), $\mathcal{M}^{\rm hid}_{4\times 4}$ is the neutralino mass matrix of the hidden sectors and is given by
\begin{eqnarray}
\mathcal{M}^{\rm hid}_{4\times 4}=
\left(\begin{matrix} 
 0 & c_1 M_1 & 0 & -c_1 s_2 M_1 \cr
 c_1 M_1 & c_1^2 m_{X_1}+s_1^2 m_1 & 0 & -s_2(s_1^2 m_1+c_1^2 m_{X_1}) \cr
0 & 0 & 0 & c_2 M_2 \cr
-c_1 s_2 M_1 & -s_2(s_1^2 m_1+c_1^2 m_{X_1})  & c_2 M_2 & c_1^2 s_2^2 m_{X_1}+c_2^2 m_{X_2}+s_1^2 s_2^2 m_1  
\end{matrix}\right),
\label{nmatrix3}
\end{eqnarray}
and $C$ contains off-diagonal elements as a result of gauge kinetic mixing between the MSSM and hidden sectors. The matrix $C$ is given by
\begin{eqnarray}
C=
\left(\begin{matrix} 
 0 & -s_1 m_1 & 0 & s_1 s_2 m_1 \cr
 0 & 0 & 0 & 0 \cr
0 & s_1 c_{\beta} s_W M_Z & 0 & -s_1 s_2 c_{\beta} s_W M_Z \cr
0 & -s_1 s_{\beta} s_W M_Z & 0 & s_1 s_2 s_{\beta} s_W M_Z   
\end{matrix}\right).
\label{nmatrix4}
\end{eqnarray}

 We label the mass eigenstates of the $8\times 8$ matrix as 
$\tilde \chi_1^0, ~\tilde \chi_2^0, ~\tilde \chi_3^0, ~\tilde \chi_4^0$ (which reside mostly in 
the visible sector) and $\tilde\xi^0_4\equiv\tilde\chi_5^0,~\tilde\xi^0_3\equiv\tilde\chi_6^0$ 
(which reside mostly in the  $U(1)_{X_1}$ sector),$~\tilde\xi^0_2\equiv\tilde\chi_7^0,~\tilde\xi^0_1\equiv\tilde\chi_8^0$
 (which reside mostly in the $U(1)_{X_2}$ sector).  
Since the mixing parameters $\delta_1$ and $\delta_2$ are very small,
the four neutralinos $\tilde\xi^0_1,\tilde\xi^0_2,\tilde\xi^0_2$ and $\tilde\xi^0_4$ reside mostly in the hidden sectors while the remaining four $\tilde \chi_i^0$
($i=1\cdots 4$) reside mostly in the MSSM sector. We will take $\tilde\xi^0_1\equiv\tilde\chi_8^0$ to be the LSP of the entire system (MSSM+hidden sectors). In the neutral gauge boson sector 
we will have mixings among the four gauge fields $X_1^\mu, X_2^\mu, B^\mu, A_3^\mu$ where
$A_3^\mu$ is the third component of the $SU(2)_L$ gauge field $A_a^\mu$ ($a=1-3$) of the Standard 
Model. After mixing and diagonalization, one obtains the vector fields $Z', Z'', Z, A_\gamma$
where $Z, A_\gamma$ are the fields of the $Z$-boson and the photon and $Z'$ and $Z''$ are the massive 
extra gauge bosons which reside mostly in the hidden sectors.

%\section{The analysis of $\tilde\chi_1^0\to \tilde\xi_1^0 \gamma$ decay width}\label{sec:loopwidth}

\section{Radiative decay of the neutralino}\label{sec:loopwidth}

Gamma ray emissions from DM decay (or annihilation) can have both galactic and extragalactic origin whose detection depends on the observation angle of a region in the sky. 
Since the DM density is highest near the center of the galaxy, one expects the strongest signal to originate from the center and so the galactic contribution would in general dominate the extragalactic component. However, the galactic center is plagued with astrophysical objects contaminating the signal making it harder to separate from the background. This can in part be diminished
 by observations at  higher latitudes.  
 There are various sources of  gamma rays. 
 Gamma ray photons can arise as a result of energetic final state radiation off of charged final state SM particles (internal Bremsstrahlung). Also, hadronic final states produce photons mainly from the decay of pions. Those photons are known as prompt and constitute a diffuse gamma ray spectrum. Another contribution to this spectrum comes from secondary sources such as inverse Compton scattering (ICS) and Bremsstrahlung. The former occurs when energetic charged particles such as electrons up-scatter photons in the interstellar radiation field (ISRF) while the latter is a result of interactions with the interstellar medium. The ISRF consists of the cosmic microwave background (CMB), thermal dust radiation and diffuse starlight. An ICS signal from DM can be hard to detect near the galactic center which is why experiments tend to analyze the spectrum near the galactic poles which becomes dominated by electrons and positrons generated by the DM halo outside the diffusion zone. Point-like sources which are mainly DM halo satellite galaxies of the Milky Way can contribute to the diffuse spectrum, but are considered weak in most cases. Subtracting the galactic diffuse emission and known point-like sources leaves us with the isotropic gamma ray background (IGRB) which is mainly made up of emissions of extragalactic origin due to blazars, quasars and star-forming galaxies, as well as active galactic nuclei. Possible galactic sources may be due to high latitude pulsars. This component of the sky proves to be important in constraining DM decay channels~\cite{Ibarra:2013cra}. Details regarding the calculations of the flux from prompt photons, ICS, antiprotons and IGRB are given in Appendix~\ref{sec:id}.

\begin{figure}[H]
 \centering
\includegraphics[width=0.32\textwidth]{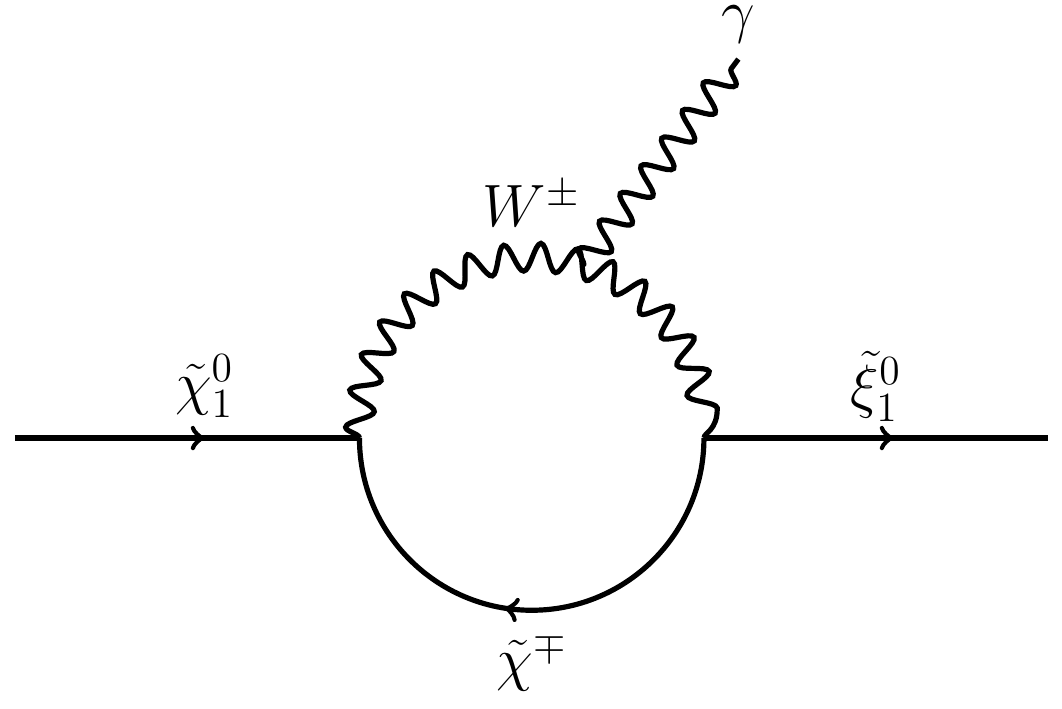}
\includegraphics[width=0.32\textwidth]{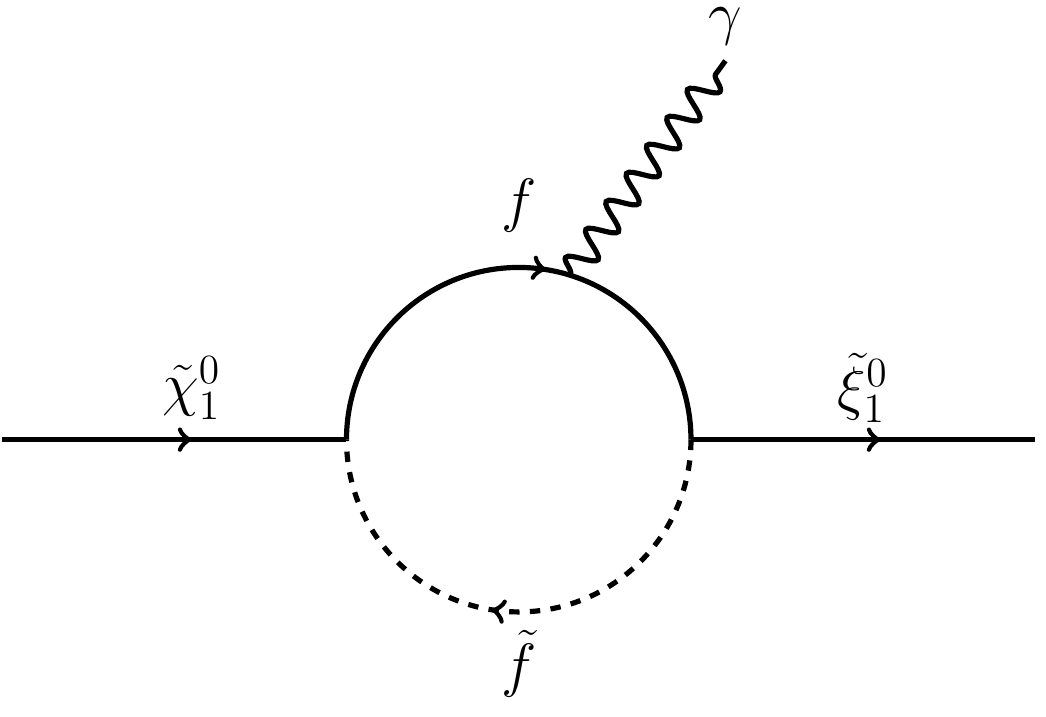}
\includegraphics[width=0.32\textwidth]{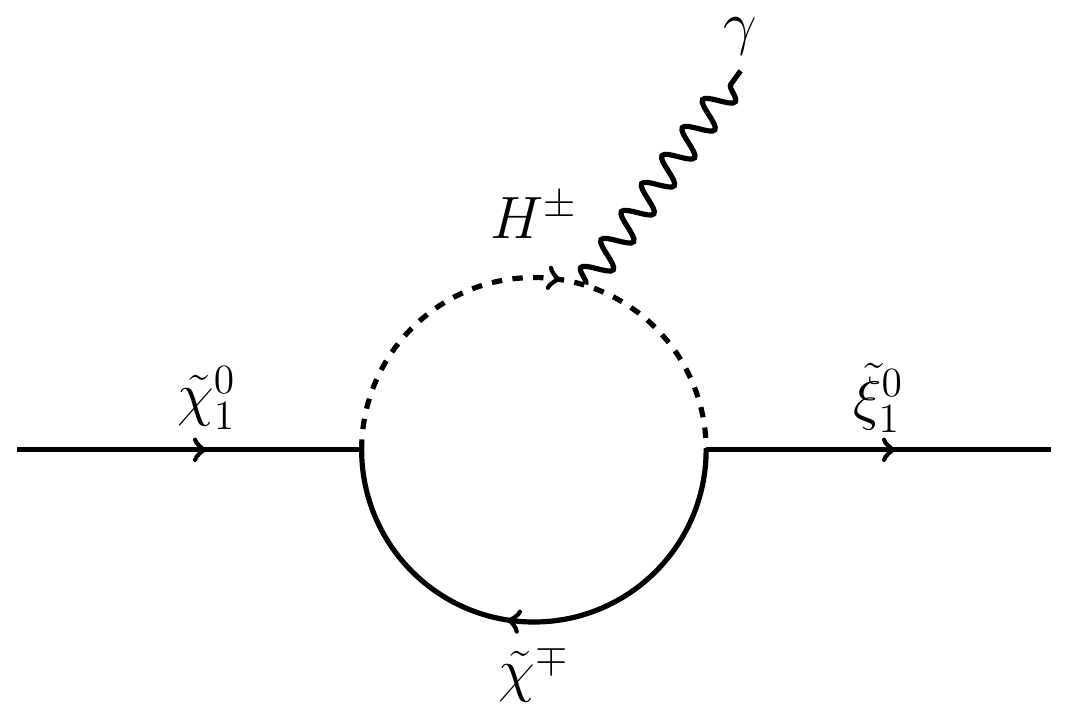} \\
\includegraphics[width=0.32\textwidth]{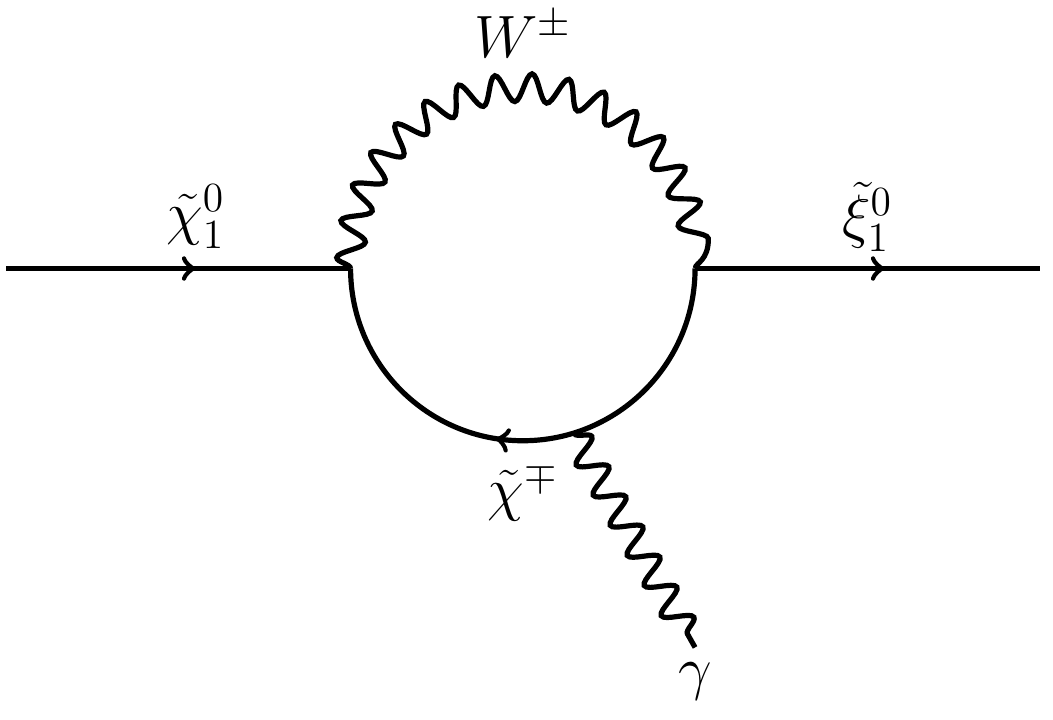}
\includegraphics[width=0.32\textwidth]{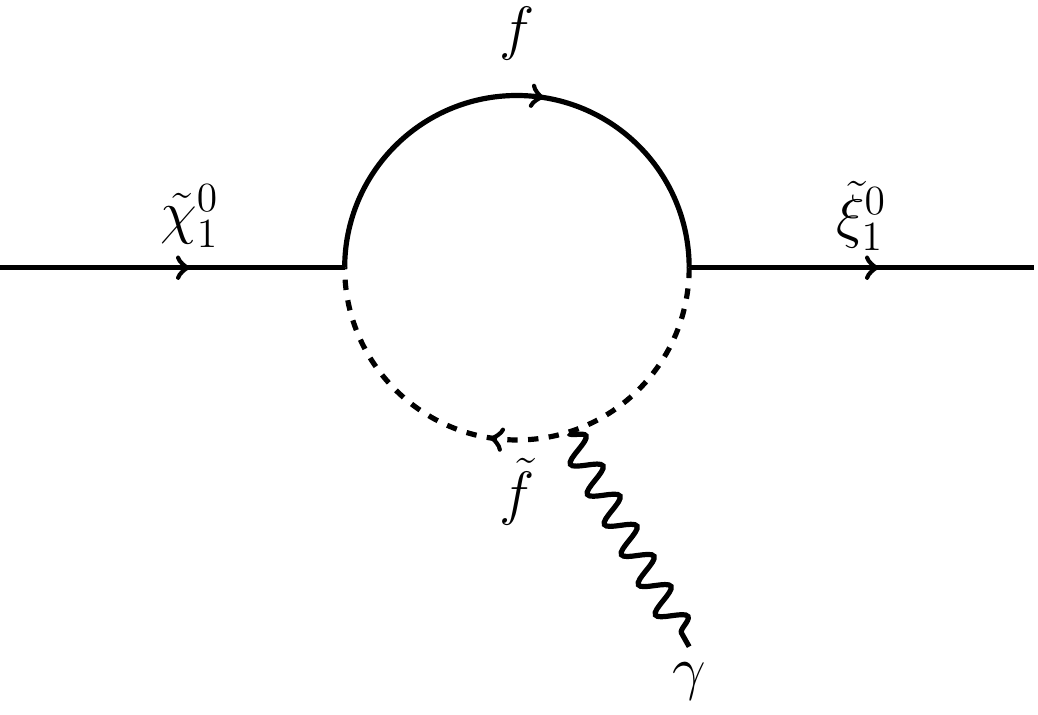}
\includegraphics[width=0.32\textwidth]{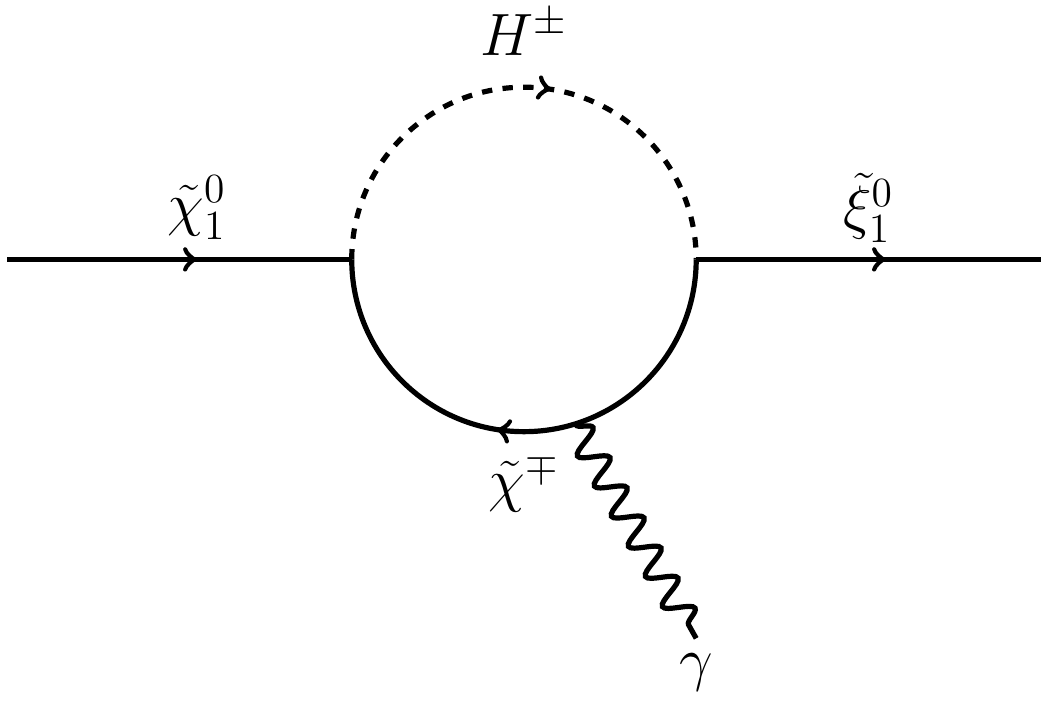}
 \caption{Loop diagrams contributing to the neutralino radiative decay $\tilde\chi_1^0\to \tilde\xi_1^0 \gamma$.}
  \label{fig1}
\end{figure}

In our model, the dominant decays of $\tilde\chi^0_1$ are three-body decays to $\tau^+\tau^-$ and a radiative loop-induced decay, $\tilde\chi_1^0\to \tilde\xi_1^0 \gamma$. 
The loop contributions to the  process $\tilde\chi_1^0\to \tilde\xi_1^0 \gamma$ are given in Fig.~\ref{fig1} which include the exchange of $W^{\pm}$$-$$\tilde\chi^{\mp}$ ($W$ boson and chargino), $f$$-$$\tilde{f}$ (fermions and sfermions) and $H^{\pm}$$-$$\tilde\chi^{\mp}$ (charged Higgs and charginos). The emission of a photon comes from either of the charged particles in the loop. The Lagrangians describing the interactions manifest in the diagrams of Fig.~\ref{fig1} are given in Appendix~\ref{sec:loop}. Using those interaction terms, we calculated the decay of width of $\tilde\chi_1^0\to \tilde\xi_1^0 \gamma$ which is given by
\begin{eqnarray}
\Gamma(\tilde\chi_1^0\to \tilde\xi_1^0 \gamma)=\frac{m^3_{\tilde\chi_1^0}}{8\pi(m_{\tilde\chi_1^0}+m_{\tilde\xi_1^0})^2}\left(1-\frac{m^2_{\tilde\xi_1^0}}{m^2_{\tilde\chi_1^0}}\right)^3[| F^{\tilde\chi_1^0\tilde\xi_1^0}_2(0)|^2+| F^{\tilde\chi_1^0\tilde\xi_1^0}_3(0)|^2],
\label{moments}
\end{eqnarray}
where the magnetic transition dipole moment form factor $F^{\tilde\chi_1^0\tilde\xi_1^0}_2(0)$ and the electric transition dipole moment form factor $F^{\tilde\chi_1^0\tilde\xi_1^0}_3(0)$ are calculated for the different contributions from the one-loop diagrams of Fig.~\ref{fig1}. 
We note in passing that while the moments vanish for Majorana particles and are in general non-vanishing for Dirac particles, the transition moments are in general non-vanishing for both 
Dirac and Majorana particles. In our case the initial and final particles are different  Majorana 
particles and we give a computation of the various contributions to the transition moments 
of Eq.~(\ref{moments}) below. Note that the computations shown below are exact and do not appear in the literature elsewhere.

\subsection{$W$ contributions}

Using the interaction of chargino, neutralino and $W$ outlined in  Eq.~(\ref{h-chargino.1}), the $W$ contributions  to the transition dipole moments are given by
\begin{align}
F^{\tilde\chi_1^0\tilde\xi_1^0}_{2W}(0)&=\sum_{
j=1}^2 g^2_2 \frac{m_{\tilde\chi^0_1}(m_{\tilde\chi^0_1}+m_{\tilde\xi^0_1})}{32\pi^2 M^2_W} (O_{8j}^L O_{1j}^{L*}+O_{8j}^R O_{1j}^{R*}-O_{1j}^L O_{8j}^{L*}-O_{1j}^R O_{8j}^{R*}) 
\tau_{1W}\left(\frac{m^2_{\tilde\chi^+_j}}{M^2_W}\right) \nonumber\\
&+\sum_{
j=1}^2 g^2_2  \frac{m_{\tilde\chi^+_j}(m_{\tilde\chi^0_1}+m_{\tilde\xi^0_1})}{32\pi^2 M^2_W} (O_{8j}^L O_{1j}^{R*}+O_{8j}^R O_{1j}^{L*}-O_{1j}^L O_{8j}^{R*}-O_{1j}^R O_{8j}^{L*})
\tau_{2W}\left(\frac{m^2_{\tilde\chi^+_j}}{M^2_W}\right),
\end{align}
and 
\begin{align}
F^{\tilde\chi_1^0\tilde\xi_1^0}_{3W}(0)=\sum_{
j=1}^2 -g^2_2  \frac{m_{\tilde\chi^+_j}(m_{\tilde\chi^0_1}+m_{\tilde\xi^0_1})}{32\pi^2 M^2_W} (O_{8j}^L O_{1j}^{R*}-O_{8j}^R O_{1j}^{L*}-O_{1j}^L O_{8j}^{R*}+O_{1j}^R O_{8j}^{L*}) 
\tau_{3W}\left(\frac{m^2_{\tilde\chi^+_j}}{M^2_W}\right),
\end{align}
where the functions $\tau_{1W}(x)$, $\tau_{2W}(x)$  and $\tau_{3W}(x)$ are given by
\begin{align}
\tau_{1W}(x)&=\frac{1}{6(x-1)^4}[-x^4 -35 x^3+39 x^2 -5x +2 +18 x^2(1+x)\ln x], \nonumber\\
\tau_{2W}(x)&=\frac{1}{(x-1)^3}[12 x^2 -12x -6 x(1+x)\ln x], \nonumber\\
\tau_{3W}(x)&=\frac{2}{(x-1)^2}\left[ -3x -\frac{3x(1+x)}{2(1-x)}\ln x\right].
\end{align}

\subsection{Squark contributions}

From the interaction of neutralino, quarks and squarks in Eq.~(\ref{neutralino-lag}), the contributions to the transition dipole moments are given by
\begin{align}
F^{\tilde\chi_1^0\tilde\xi_1^0}_{2\tilde{q}}(0)&=-\sum_{
i=1}^2 Q_q \frac{(m_{\tilde\chi^0_1}+m_{\tilde\xi^0_1})}{64\pi^2 m_{q}} (M_{qi8} K_{qi1}^{*}+K_{qi8}  M_{qi1}^{*}-M_{qi1} K_{qi8}^{*}-K_{qi1}  M_{qi8}^{*}) 
\tau_{1\tilde{q}}\left(\frac{m^2_{\tilde{q}_i}}{m^2_{q}}\right) \nonumber\\
&-\sum_{
i=1}^2 Q_q \frac{m_{\tilde\chi^0_1}(m_{\tilde\chi^0_1}+m_{\tilde\xi^0_1})}{192\pi^2 m^2_{q}} 
 (M_{qi8} M_{qi1}^{*}+K_{qi8} K_{qi1}^{*}-M_{qi1} M_{qi8}^{*}-K_{qi1} K_{qi8}^{*})
\tau_{2\tilde{q}}\left(\frac{m^2_{\tilde{q}_i}}{m^2_{q}}\right),
\end{align}
and
\begin{equation}
F^{\tilde\chi_1^0\tilde\xi_1^0}_{3\tilde{q}}(0)=-\sum_{i=1}^2
 Q_q \frac{m_q (m_{\tilde\chi^0_1}+m_{\tilde\xi^0_1})}{32\pi^2 m^2_{\tilde{q}_i}} (M_{qi1} K_{qi8}^{*}-K_{qi1}  M_{qi8}^{*}-M_{qi8} K_{qi1}^{*}+K_{qi8}  M_{qi1}^{*}) 
\tau_{3\tilde{q}}\left(\frac{m^2_{\tilde{q}_i}}{m^2_{q}}\right),
\end{equation}
where the form factors $\tau_{i\tilde{q}}(x)$ are given by
\begin{align}
\tau_{1\tilde{q}}(x)&=\frac{1}{(x-1)^3}[ -4x^2 +4x  +2 x(1+x)\ln x],\nonumber\\
\tau_{2\tilde{q}}(x)&=\frac{3}{(x-1)^4}[-x^3+ x^2 +x-1 -2 x(1-x)\ln x],\nonumber\\
\tau_{3\tilde{q}}(x)&=\frac{1}{(x-1)^2}[ x -1-\ln x],
\label{qform}
\end{align}
and $Q_q$ is the charge of the quark involved in the loop. All quark flavors are included in the calculations. 

\subsection{Slepton contributions}
From the interaction of slepton-lepton-neutralinos, Eq.~(\ref{slepton-lepton}),
 the slepton contributions to the transition dipole moments are given by
\begin{align}
F^{\tilde\chi_1^0\tilde\xi_1^0}_{2\tilde{\ell}}(0)&=\sum_{
i=1}^2  \frac{(m_{\tilde\chi^0_1}+m_{\tilde\xi^0_1})}{64\pi^2 m_{\ell}} (F_{8i} Z_{1i}^{*}+Z_{8i}  F_{1i}^{*}-F_{1i} Z_{8i}^{*}-Z_{1i}  F_{8i}^{*}) 
\tau_{1\tilde{\ell}}\left(\frac{m^2_{\tilde{\ell}_i}}{m^2_{\ell}}\right) \nonumber\\
&+\sum_{
i=1}^2  \frac{m_{\tilde\chi^0_1}(m_{\tilde\chi^0_1}+m_{\tilde\xi^0_1})}{192\pi^2 m^2_{\ell}} 
 (F_{8i} F_{1i}^{*}+Z_{8i}  Z_{1i}^{*}-F_{1i} F_{8i}^{*}-Z_{1i}  Z_{8i}^{*})
\tau_{2\tilde{\ell}}\left(\frac{m^2_{\tilde{\ell}_i}}{m^2_{\ell}}\right),
\end{align}
and
\begin{align}
F^{\tilde\chi_1^0\tilde\xi_1^0}_{3\tilde{\ell}}(0)=\sum_{i=1}^2
  \frac{m_{\ell} (m_{\tilde\chi^0_1}+m_{\tilde\xi^0_1})}{32\pi^2 m^2_{\tilde{\ell}_i}} (F_{1i} Z_{8i}^{*}-Z_{1i}  F_{8i}^{*}-F_{8i} Z_{1i}^{*}+Z_{8i}  F_{1i}^{*}) 
\tau_{3\tilde{\ell}}\left(\frac{m^2_{\tilde{\ell}_i}}{m^2_{\ell}}\right),
\end{align}
where the form factors $\tau_{i\tilde{\ell}}(x)$ have the same form as $\tau_{i\tilde{q}}(x)$ given in Eq.~(\ref{qform}). Contributions from electrons, muons and taus are included. 

\subsection{Charged Higgs contributions}

Using the interaction of charged Higgs, neutralino and chargino, Eq.  (\ref{h-charge-neu.1}), the charged Higgs contributions to the transition dipole moments are given by
\begin{align}
F^{\tilde\chi_1^0\tilde\xi_1^0}_{2H}(0)&=\sum_{
j=1}^2 \frac{(m_{\tilde\chi^0_1}+m_{\tilde\xi^0_1})}{64\pi^2 m_{\tilde\chi^+_j}} (\eta_{8j}^L \eta_{1j}^{R*}+\eta_{8j}^R \eta_{1j}^{L*}-\eta_{1j}^L \eta_{8j}^{R*}-\eta_{1j}^R \eta_{8j}^{L*})
\tau_{1H}\left(\frac{m^2_{H^-}}{m^2_{\tilde\chi^+_j}}\right),\nonumber\\
&+\sum_{
j=1}^2 \frac{m_{\tilde\chi^0_1}(m_{\tilde\chi^0_1}+m_{\tilde\xi^0_1})}{192\pi^2 m^2_{\tilde\chi^+_j}} (\eta_{8j}^L \eta_{1j}^{L*}+\eta_{8j}^R \eta_{1j}^{R*}-\eta_{1j}^L \eta_{8j}^{L*}-\eta_{1j}^R \eta_{8j}^{R*}) 
\tau_{2H}\left(\frac{m^2_{H^-}}{m^2_{\tilde\chi^+_j}}\right).
\end{align}
and 
\begin{equation}
F^{\tilde\chi_1^0\tilde\xi_1^0}_{3H}(0)=\sum_{
j=1}^2 \frac{m_{\tilde\chi_j^+}(m_{\tilde\chi^0_1}+m_{\tilde\xi^0_1})}{32\pi^2 m^2_{H^+}} (\eta_{1j}^L \eta_{8j}^{R*}-\eta_{1j}^R \eta_{8j}^{L*}-\eta_{8j}^L \eta_{1j}^{R*}+\eta_{8j}^R \eta_{1j}^{L*}) 
\tau_{3H}\left(\frac{m^2_{\tilde\chi^+_j}}{m^2_{H^-}}\right).
\end{equation}
The couplings $\eta^L_{Aj}$ and $\eta^R_{Aj}$ are related to the quantities defined in Eq.~(\ref{spcoupling}) as follows
\begin{equation}
\begin{aligned}
\eta^L_{Aj}&=\alpha^S_{Aj}-\alpha^P_{Aj}, \\
\eta^R_{Aj}&=\alpha^S_{Aj}+\alpha^P_{Aj},
\end{aligned}
\end{equation}
and the functions  $\tau_{iH}(x)$ are the same as $\tau_{i\tilde{q}}(x)$ of Eq.~(\ref{qform}).

\section{Numerical analysis}\label{sec:numerics}

The visible sector (the MSSM/SUGRA) and the hidden sectors of the model described in section~\ref{sec:model} are essentially decoupled, so one can use the MSSM/SUGRA implemented in the spectrum generator \code{SPheno-4.0.4}~\cite{Porod:2003um,Porod:2011nf} which runs the two-loop renormalization group equations (RGE) starting from a high scale input and taking into account threshold effects to produce the loop-corrected sparticle masses and calculate their decay widths. To determine the dark matter relic density and the DM thermally averaged annihilation cross-section we use \code{micrOMEGAs-5.2.1}~\cite{Belanger:2014vza}. Note that one can also take the parameters of the extended sector at the GUT scale and run them down to the low scale using new model implementations in \code{SPheno} and \code{micrOMEGAs}. However, we have shown in previous works (see e.g.~\cite{Aboubrahim:2019qpc}) that the running of such parameters is very mild and barely changes since the beta functions are proportional to the gauge kinetic mixing coefficients.
The input parameters of the $U(1)_X$-extended MSSM/SUGRA~\cite{mSUGRA} are of the usual non-universal SUGRA model with additional parameters (all at the GUT scale) as defined in section~\ref{sec:model}:
$m_0$, $A_0$, $m_1$, $m_2$, $m_3$, $M_1$, $M_2$, $m_{X_1}$, $m_{X_2}$, $\delta_1$, $\delta_2$, $\tan\beta$, $\text{sgn}(\mu)$. We run a scan of the parameter space while retaining points which satisfy the Higgs boson mass constraint at $125\pm 2$ GeV and the DM relic density as reported by the Planck experiment~\cite{Aghanim:2018eyx}
\begin{equation}
\Omega h^2=0.1198\pm 0.0012.
\label{relic}
\end{equation}
The remaining points are further filtered after imposing LHC constraints on the electroweakino masses and bounds on proton-neutralino spin-independent cross-section from XENON1T~\cite{Aprile:2018dbl}. We select points that have relatively light stau masses, $\mathcal{O}$(1$-$2) TeV, and the lightest neutralino of the hidden sector, $\tilde\xi^0_1$ as the LSP and the MSSM/SUGRA neutralino $\tilde\chi^0_1$ as the NLSP. The second hidden sector neutralino is much heavier. Ten benchmarks are selected for this analysis and their high scale input parameters are shown in Table~\ref{tab1}. 

\begin{table}[H]
\begin{center}
\begin{tabular}{l|ccccccccc}
\hline\hline\rule{0pt}{3ex}
Model & $m_0$ & $A_0$ & $m_1$ & $m_2$ & $m_3$ & $M_2$ & $m_{X_2}$ & $\tan\beta$ & $\delta=\delta_1\delta_2$ \\
\hline\rule{0pt}{3ex}  
\!\!(a) & 2497 & 7916 & 788 & 523 & 6696 & 750 & 3500 & 37 & $1.80\times 10^{-23}$  \\
(b) & 4913 & -13831 & 1066 & 650 & 7743 & 400 & 1200 & 46 & $1.05\times 10^{-23}$ \\
(c) & 1616 & 7988 & 1302 & 788 & 6368 & 250 & 950 & 25 & $2.73\times 10^{-24}$ \\
(d) & 3177 & -9436 & 1415 & 787 & 2883 & 450 & 1050 & 40 & $1.90\times 10^{-24}$ \\
(e) & 1298 & -11473 & 1793 & 997 & 3502 & 800 & 4700 & 12 &  $5.10\times 10^{-25}$ \\
(f) & 1475 & -8356 & 2076 & 1276 & 5868 & 700 & 4900 & 20 & $4.55\times 10^{-24}$ \\
(g) & 2078 & -9650 & 2265 & 1244 & 4114 & 650 & 5000 & 25 &  $2.00\times 10^{-25}$ \\
(h) & 2195 & -14006 & 3025 & 1629 & 4076 & 550 & 5100 & 18 & $1.00\times 10^{-24}$ \\
(i) & 1334 & -6155 & 3547 & 1927 & 4836 & 600 & 5500 & 10 & $6.00\times 10^{-26}$ \\
(j) & 1889 & -20595 & 3977 & 2180 & 6428 & 750 & 7500 & 8 & $1.00\times 10^{-26}$ \\
\hline
\end{tabular}\end{center}
\caption{Input parameters for the benchmarks used in this analysis (where all masses are in GeV) and
 where we set $M_1=50$ TeV and $m_{X_1}=100$ GeV.}
\label{tab1}
\end{table}

The choice of the parameters in Table~\ref{tab1} is such that the hidden sector neutralinos 
$\tilde\xi^0_2$, $\tilde\xi^0_3$, $\tilde\xi^0_4$ are very heavy and $\tilde\xi^0_1$ is the lightest 
with a mass smaller than that of $\tilde\chi_1^0$ which is the LSP of the visible sector.  The kinetic
mixing of the visible and hidden sector $U(1)$'s in large volume compactifications can be $\mathcal{O}(10^{-12})$.
However, under special circumstances where the mixings arise from non-perturbative contributions
suppressed by factor of $e^{-aT}$, where $T$ is some modulus, the mixings can be much smaller
and lie in the range $10^{-20}-10^{-26}$~\cite{Goodsell:2009xc}.
 In the analysis of section~\ref{sec:model}, we assume that the kinetic
mixings are of the generic type in large volume compactifications and one can generate much 
smaller couplings in the range $10^{-20}- 10^{-26}$ by considering products of two $U(1)'$ as discussed
in section~\ref{sec:model}. Of course, one can also simply assume the ultra small coupling directly but the analysis of 
section~\ref{sec:model} gives an alternative way to generate very small mixings. The smallness of the gauge kinetic mixing parameter $\delta\sim 10^{-24}$ ensures that the decay 
$\tilde\chi^0_1\to\tilde\xi^0_1\gamma$ is 
long-lived with a lifetime greater than the age of the universe. Because of such suppressed decays, $\tilde\xi^0_1$ makes a negligible part of the relic density and for all practical 
purposes  the dark matter relic density 
is entirely due to $\tilde\chi^0_1$. This is so because while a relic density for feeble particles can be generated by the freeze-in mechanism~\cite{Hall:2009bx} (for recent works see~\cite{Aboubrahim:2019kpb,Aboubrahim:2020wah,Aboubrahim:2020lnr}), their  ultraweak couplings to the Standard Model particles makes the production extremely small.

The relevant part of the mass spectrum arising from the benchmarks of Table~\ref{tab1}
is shown in Table~\ref{tab2} along with the DM relic density. Here we note that a light stau mass can be achieved by choosing a small $m_0$ and a large $A_0$ to  split the stau masses.  

\begin{table}[H]
\begin{center}
\begin{tabular}{l|ccccccccc}
\hline\hline\rule{0pt}{3ex}
Model  & $h^0$ & $\mu$ & $\tilde\chi_1^0$ & $\tilde\chi_1^\pm$ & $\tilde{\tau}$ & $\tilde{\xi}^0_1$ & $\tilde t$ & $\tilde g$ & $\Omega h^2$  \\
\hline\rule{0pt}{3ex} 
\!\!(a) & 124.0 & 5899 & 301 & 325 & 937 & 154 & 9901 & 13014 & 0.115  \\
(b) & 124.1 & 9821 & 451 & 479 & 860 & 121 & 10106 & 15094 & 0.119  \\
(c) & 124.0 & 5820 & 538 & 566 & 713 & 62 & 9336 & 12440 &  0.110  \\
(d) & 126.5 & 5034 & 627 & 656 & 936 & 166 & 3437 & 6019 &  0.117  \\
(e) & 125.5 & 6373 & 789 & 815 & 942 & 132 & 3135 & 7073 &  0.119 \\
(f) & 126.4 & 7324 & 911 & 1027 & 915 & 98 & 7648 & 11491 &  0.116 \\
(g) & 126.9 & 6216 & 1007 & 1026 & 1054 & 83 & 4927 & 8252 & 0.117  \\
(h) & 125.2 & 7529 & 1361 & 1363 & 1371 & 59 & 3368 & 8164 & 0.121  \\
(i) & 125.7 & 6013 & 1590 & 1591 & 1757 & 65 & 6429 & 9565 & 0.123  \\
(j) & 124.8 & 11340 & 1804 & 1816 & 1975 & 74 & 5147 & 12469 & 0.124 \\
\hline
\end{tabular}\end{center}
\caption{For the benchmarks  of Table~\ref{tab1}, a display of the Higgs boson ($h^0$) mass, the $\mu$ parameter, and the sparticle spectrum mass consisting of the electroweakinos, the stau,  the hidden sector neutralino, the stop and gluino, respectively, computed at the electroweak scale. Also shown is the DM relic density. All masses are in GeV.}
\label{tab2}
\end{table}

It is found that all the neutralinos for the  benchmarks of Table~\ref{tab1}
are bino-like. This makes satisfying the relic density harder which is why a coannihilating partner is needed. The chargino plays that role and sits just above the NLSP mass. This so-called compressed spectrum is less constrained by LHC experiments. However, several searches in this regime have been carried out in the 2-lepton and 3-lepton final states~\cite{Aaboud:2018sua,Aad:2019qnd,ATLAS:2020ckz} with the most stringent limits set by Refs.~\cite{Aad:2015eda,Aaboud:2018jiw}. A neutralino with mass less than 260 GeV and an NLSP chargino of mass less than 300 GeV are excluded. The electroweakino spectrum in our analysis is safely outside the exclusion limits.  The stops and gluinos are very heavy, the result of choosing a large $m_3$ which is crucial for getting the correct Higgs mass in light of the small scalar mass $m_0$. The lightness of the stau and the largeness of $\mu$ makes the two-body decay channels $\tilde\chi^0_1\to \tilde\xi^0_1 Z$ and $\tilde\chi^0_1\to \tilde\xi^0_1 h$ suppressed in comparison with the three-body decay channel $\tilde\chi^0_1\to \tau^+\tau^-\tilde\xi^0_1$. Furthermore, the DM annihilation will predominantly proceed via a $t$-channel stau making the process $\tilde\chi^0_1\tilde\chi^0_1\to\tau^+\tau^-$ the dominant one.

\begin{table}[H]
\begin{center}
\begin{tabular}{l|cccc}
\hline\hline\rule{0pt}{3ex}
Model  & $\langle\sigma v\rangle_{\tau\tau}$ & $\Gamma(\tilde\chi^0_1\rightarrow\tilde\xi^0_1\tau^+\tau^-)$ & $\Gamma(\tilde\chi^0_1\rightarrow\tilde\xi^0_1\gamma)$ & $E_{\gamma}$ \\
\hline\rule{0pt}{3ex} 
\!\!(a) & $5.2\times 10^{-30}$ & $8.2\times 10^{-54}$ & $2.7\times 10^{-54}$ & 111.1 \\
(b) & $4.1\times 10^{-30}$ & $3.6\times 10^{-52}$ & $1.6\times 10^{-54}$ & 209.3 \\
(c) & $4.0\times 10^{-29}$ & $6.9\times 10^{-52}$ & $1.8\times 10^{-54}$ & 265.4 \\
(d) & $3.2\times 10^{-30}$ & $4.9\times 10^{-54}$ & $1.7\times 10^{-54}$ & 291.5 \\
(e) & $7.6\times 10^{-30}$ & $4.7\times 10^{-53}$ & $1.9\times 10^{-54}$ & 383.5 \\
(f) & $1.6\times 10^{-29}$ & $7.4\times 10^{-51}$  & $2.1\times 10^{-54}$ & 450.2 \\
(g) & $5.7\times 10^{-30}$ & $3.4\times 10^{-51}$ & $2.2\times 10^{-53}$ & 500.1 \\
(h) & $4.3\times 10^{-30}$ & $2.6\times 10^{-51}$ & $1.5\times 10^{-52}$ & 679.2 \\
(i) & $2.0\times 10^{-28}$ & $5.2\times 10^{-53}$ & $1.0\times 10^{-54}$ & 793.7 \\
(j) & $3.6\times 10^{-31}$ & $9.1\times 10^{-54}$ & $4.8\times 10^{-55}$ & 900.5 \\
\hline
\end{tabular}\end{center}
\caption{For the benchmarks of Table~\ref{tab1}, a display of the DM annihilation cross-section, the radiative decay width and three-body decay width of $\tilde\chi^0_1$ and the attendant
photon energy $E_\gamma$. The decay width and the photon energy are in GeV and the thermally averaged cross-section is in units of cm$^3$/s.}
\label{tab3}
\end{table}

We present in Table~\ref{tab3} the DM annihilation cross-section into $\tau^+\tau^-$ final state. Fermi-LAT and MAGIC collaborations~\cite{Ahnen:2016qkx} have set limits on this cross-section excluding values above $\mathcal{O}(10^{-26}-10^{-25})$ cm$^3$/s in the range 100 GeV to 1 TeV. Our cross-sections are far below this limit and thus not excluded. We also display the three-body decay width and the radiative decay width of $\tilde\chi^0_1$ determined using Eq.~(\ref{moments}) as well as the photon energy using Eq.~(\ref{egdecay}).  The three-body decay width is larger than the radiative decay width in most of the benchmarks and are comparable in some. This result has been observed before~\cite{Garny:2010eg} where the two neutralinos $\tilde\chi^0_1$ and $\tilde\xi^0_1$ have opposite CP phases. 

\begin{figure}[H]
    \includegraphics[width=0.49\textwidth]{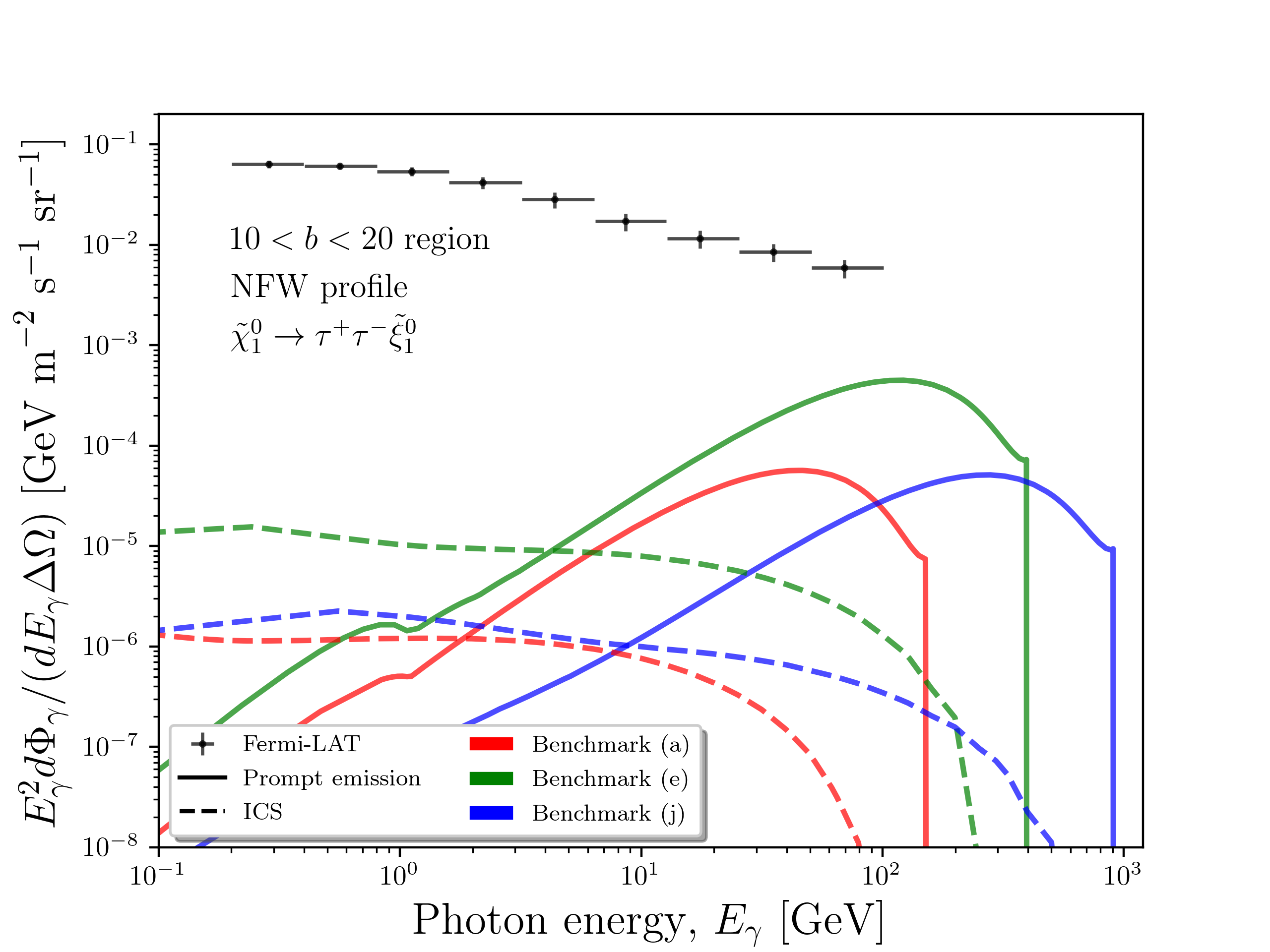}
     \includegraphics[width=0.49\textwidth]{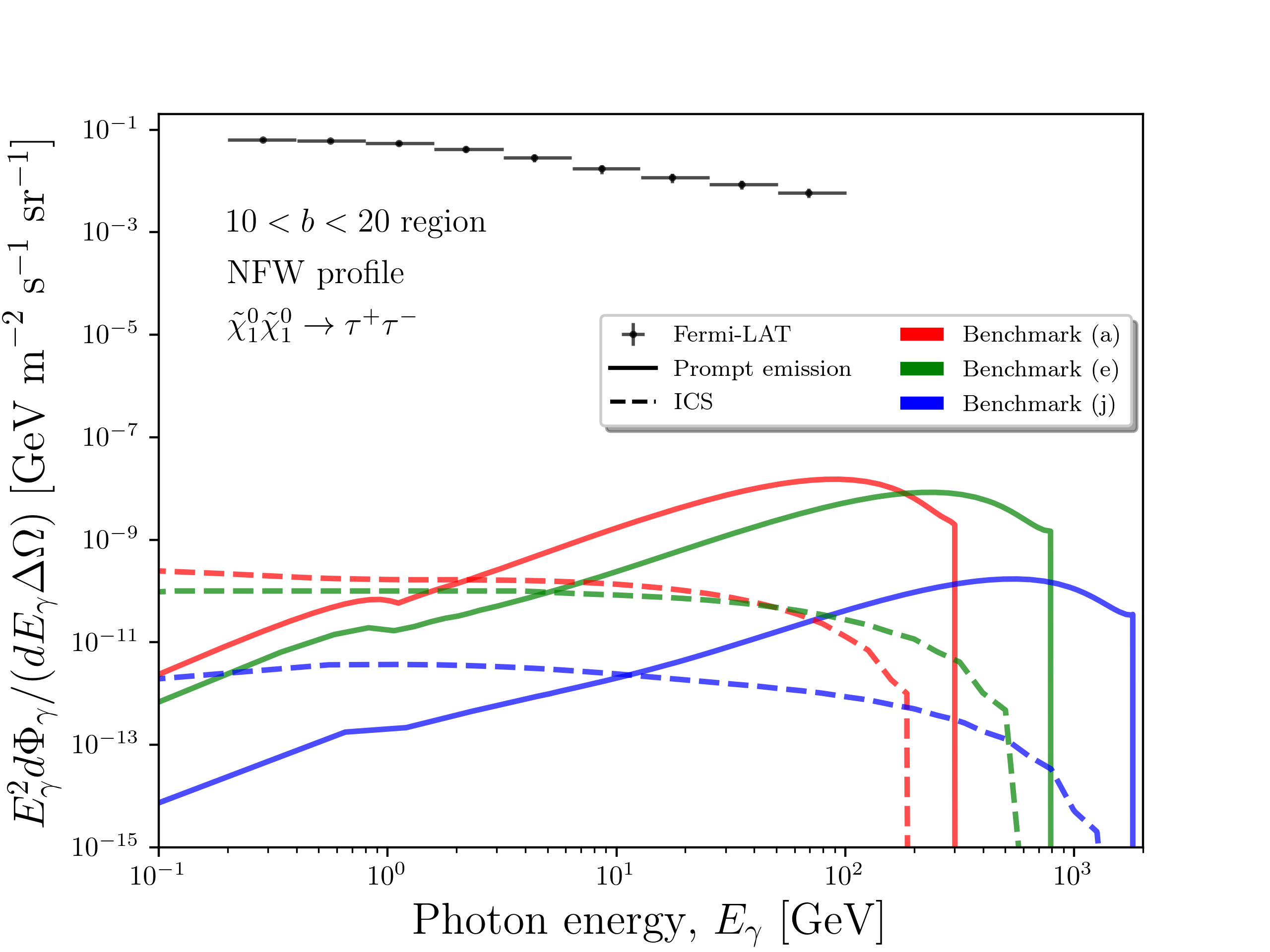}
   \caption{The gamma ray flux from the decay (left panel) and annihilation (right panel) of
   $\tilde \chi_1^0$ for three benchmarks of Table~\ref{tab1} due to prompt emission and ICS. Also shown are the Fermi-LAT data points.}
	\label{fig2}
\end{figure}

We first examine the indirect detection signals from the three-body decay of $\tilde\chi^0_1$. The three-body decay width is calculated using standard formalism~\cite{Baer:2006rs} and making the appropriate replacements of couplings to account for the hidden sector neutralino. As noted, 
the three-body decay of $\tilde\chi^0_1$ is dominated by $\tau^+\tau^-$ in the final state which 
eventually decay into electrons where the electrons radiate photons via FSR. Hadronic decays of the taus can also produce photons mainly from the decay of pions. A similar outcome is expected from the annihilation channel. All of these processes are the origin of prompt photons whose flux is determined in the region $10<b<20$ of the sky using Eqs.~(\ref{sdecay}) and~(\ref{sann}). Also, the gamma ray flux of secondary photons due to ICS is calculated using Eqs.~(\ref{icsdecay}) and~(\ref{icsann}) in the same region. We show in Fig.~\ref{fig2} the resulting flux of prompt and ICS photons versus the photon energy for the decay (left panel) and the annihilation  (right panel) for three benchmarks of Table~\ref{tab1}.   

It is seen in Fig.~\ref{fig2} that ICS dominates over prompt photons for small energies while prompt photons have higher intensities for higher energies which is consistent with a photon spectrum obtained from pion decays. Also plotted are the Fermi-LAT data points in the same region of observation. For both cases of decay and annihilation, the sum of prompt and secondary photons' flux is below the observed gamma ray flux. 

\begin{figure}[H]
   \includegraphics[width=0.49\textwidth]{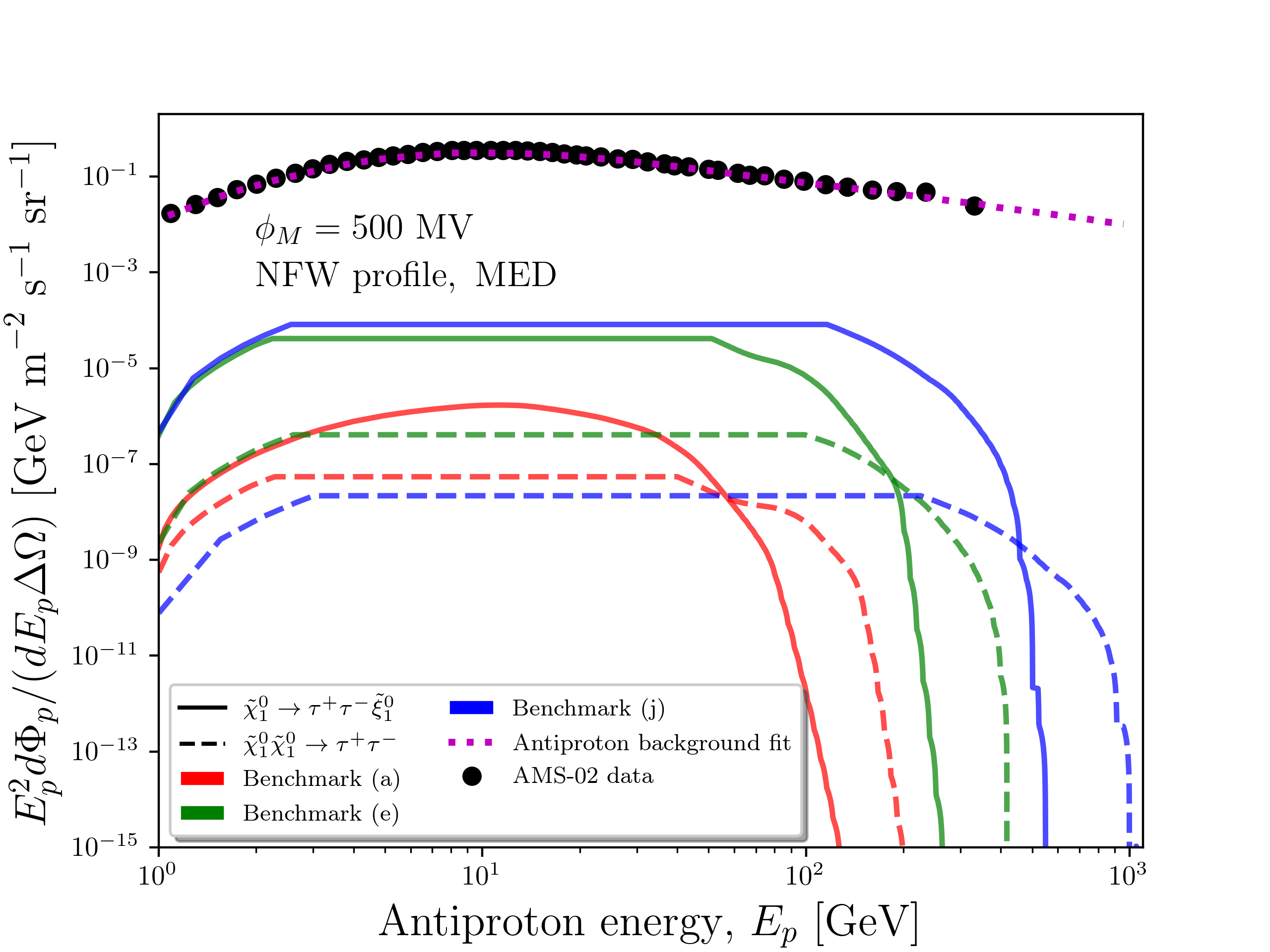}
   \includegraphics[width=0.49\textwidth]{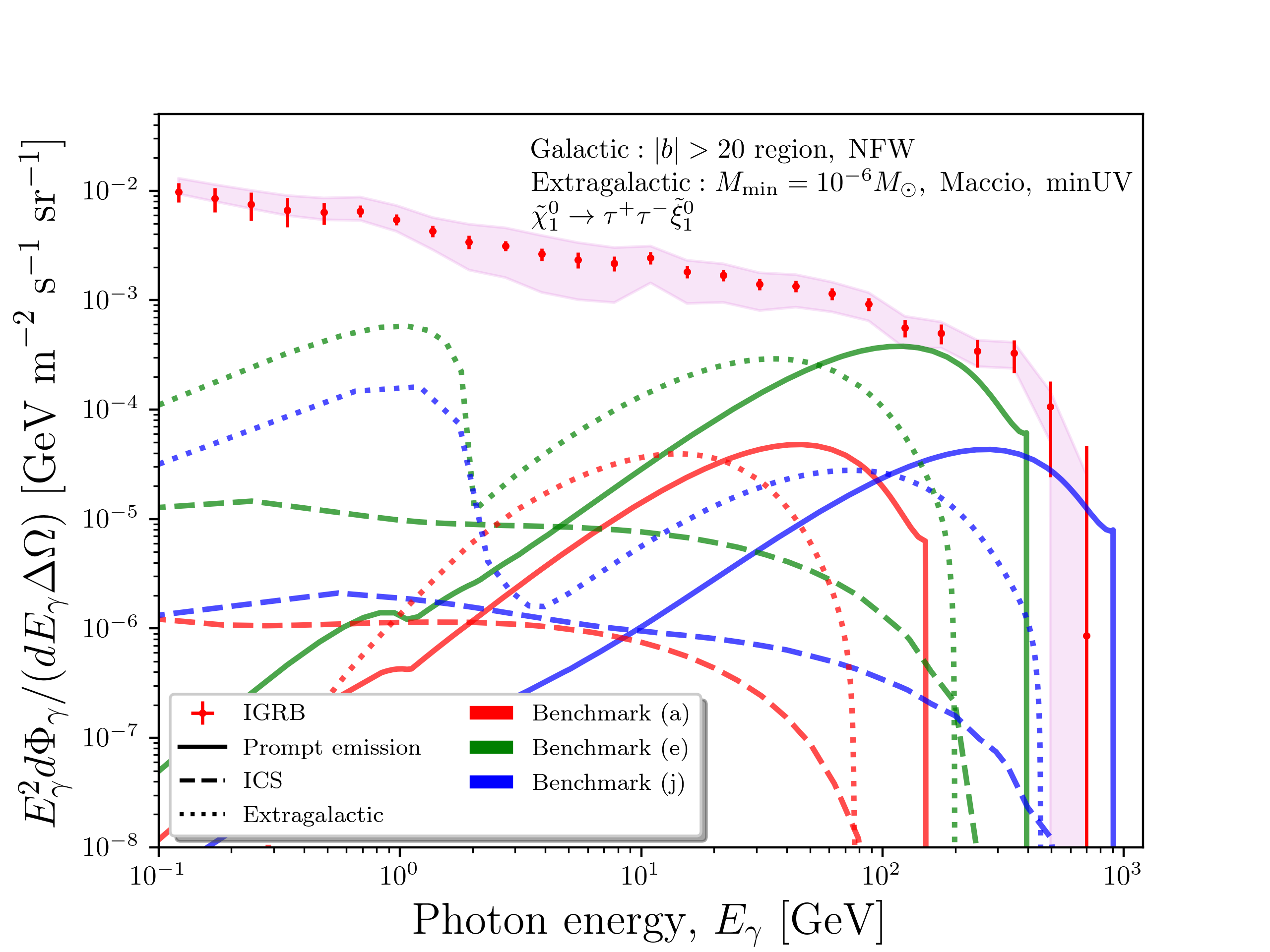}
   \caption{Left panel: Contributions from the annihilation and decay of the MSSM neutralino $\tilde\chi^0_1$ to the antiproton flux for three of the benchmarks of Table~\ref{tab1}. The antiproton background and AMS data are also shown. Right panel: the gamma ray flux from the IGRB (for decay only) showing contributions from prompt photons, ICS and extragalatic photons. Fermi-LAT data points are in red with error bars and the uncertainty band from foreground modeling. }
	\label{fig3}
\end{figure}

Since taus can decay hadronically, it is important to make sure that the antiproton flux from decay and annihilation of DM does not exceed the antiproton background. This background fits well the estimates from known astrophysical processes and allows very little room for additional hadronic decays or annihilation of DM.
Despite the reported excesses which were discussed in the introduction, uncertainties in the DM density profiles, the propagation parameters used in solving the antiproton transport equation and solar modulation can wash away the excess at low energy while the higher energy excess can be argued away due to astrophysical processes. The antiproton flux for three benchmarks of Table~\ref{tab1} is calculated using Eqs.~(\ref{antipa}) and~(\ref{mod}) and plotted against the antiproton energy in the left panel of Fig.~\ref{fig3}. The recent AMS-02 data~\cite{Aguilar:2016kjl} is shown along the estimated background fit.

The antiproton fluxes coming from the decay and annihilation of DM for the benchmarks of
Table~\ref{tab1} lie well below the observed background. Another possible antimatter channel is that of positrons. Since the leptonic decays of taus are smaller than the hadronic ones, one can safely assume that our benchmarks do not contribute to the observed positron excess and so is not constraining for 
the model. 
In the right panel of Fig.~\ref{fig3} we show contributions from DM decay to the IGRB for three benchmarks of Table~\ref{tab1}. Photon emission from prompt decay (solid line) and ICS (dashed line) as well as the extragalactic component (dotted line) lie below the diffuse IGRB measured by Fermi-LAT~\cite{Ackermann:2014usa} and shown as red points with error bars due to systematic and statistical uncertainties. The magenta band corresponds to uncertainties on IGRB from foreground modeling. The case of annihilation is not shown since it is much smaller than the decay case and therefore can be neglected.

There are several experiments planned or are underway on indirect detection of decaying DM signatures. One of these is the Square Kilometer Array (SKA)~\cite{Bull:2018lat} which can reach high frequencies and would be a suitable probe of the high DM masses. SKA is expected to have improved efficiency in separating foregrounds thanks to its intercontinental baseline length and large effective area. In regards to the DM decay lifetime, SKA can reach 2 to 3 orders of magnitude in sensitivity higher than Fermi-LAT~\cite{Ghosh:2020ipv} for DM masses up to 100 TeV. 
Another telescope array known as Cherenkov Telescope Array (CTA)~\cite{Abdalla:2020gea} is being built with more than 100 telescopes in the Northern and Southern hemispheres. CTA will scan the skies looking for high energy gamma rays and will be 10 times more sensitive than H.E.S.S.~\cite{Aharonian:2008aa,Aharonian:2009ah}, MAGIC~\cite{Ahnen:2017pqx,Ahnen:2016qkx} and VERITAS~\cite{Archambault:2017wyh,Zitzer:2015eqa}.

\begin{figure}[H]
\centering
 \includegraphics[width=0.49\textwidth]{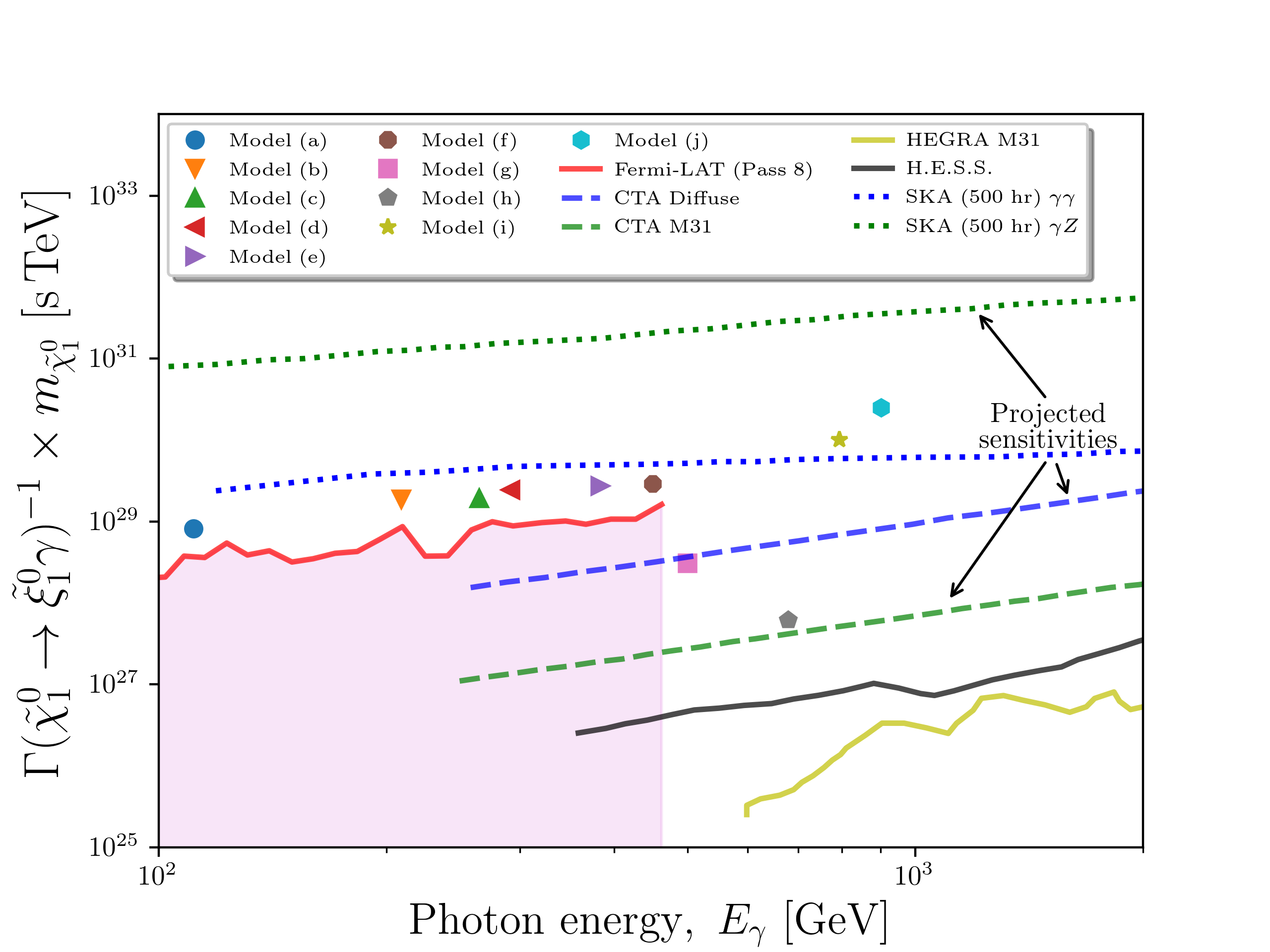}
\includegraphics[width=0.49\textwidth]{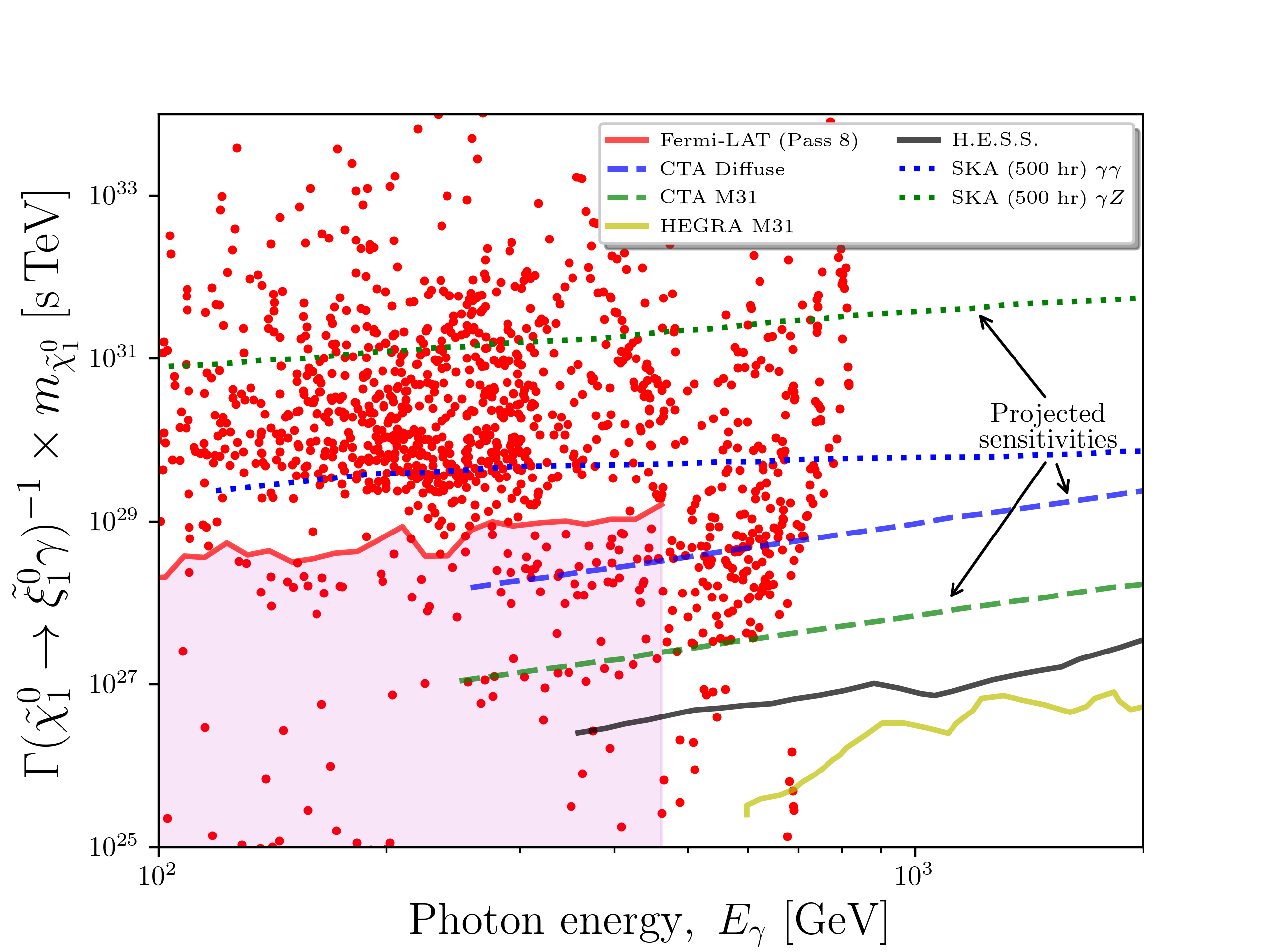}
   \caption{Left panel: A plot of $m_{\tilde\chi^0_1}/\Gamma(\tilde\chi^0_1\rightarrow\tilde\xi^0_1\gamma)$, in TeV s, versus the photon energy in GeV 
   for a set of ten benchmarks consistent with all collider and astrophysical constraints. 
   The model points are accessible in future improved experiments. Right panel: A scatter plot in the same variables shows that the parameter space in which a decaying neutralino consistent
   with all the current constraints can reside is quite large and a significant part of the parameter space
   will be accessible to more sensitive future experiments such as SKA and CTA~\cite{Garny:2010eg,Consortium:2010bc}. The ranges of the scanned parameters are: $800~\text{GeV}<m_0<5~\text{TeV}$, $-5<A_0/m_0<5$, $700~\text{GeV}<m_1<5.5~\text{TeV}$, $400~\text{GeV}<m_2<4~\text{TeV}$, $2~\text{TeV}<m_3<6.5~\text{TeV}$, $1<\tan\beta<50$, $500~\text{GeV}<M_1<5~\text{TeV}$ and $m_{X_2}=6$ TeV.
The current and future limits from dark matter indirect detection experiments are 
shown in both panels. }
	\label{fig4}
\end{figure}

Since the  benchmarks of Table~\ref{tab1} satisfy all the indirect detection constraints, 
we can now use them to  investigate the possibility of observing a sharp monochromatic spectral line  
 in the decay $\tilde\chi^0_1\to\tilde\xi^0_1\gamma$. In Fig.~\ref{fig4}, we exhibit $m_{\tilde\chi^0_1}/\Gamma(\tilde\chi^0_1\rightarrow\tilde\xi^0_1\gamma)$ plotted against the photon energy for the  benchmarks of Table~\ref{tab1} (left panel). We also exhibit the current limits from Fermi-LAT~\cite{Abdo:2010nc,Abdo:2010nz}, HEGRA~\cite{Aharonian:2003xh} and H.E.S.S.~\cite{Aharonian:2008aa,Aharonian:2009ah} as well as the projections from the future CTA and SKA experiments.  The limits from SKA are derived from the $\gamma\gamma$ and $\gamma Z$ channels assuming 500 hours of operation. One can notice that the limit from $\gamma Z$ is more stringent since $Z$ fragments into visible SM particles. Here one finds that none of the model points of Table~\ref{tab1} are yet
 excluded by experiment. Thus benchmarks (a)$-$(f) lie just above the current Fermi-LAT limit and can be probed with higher sensitivities while the rest of the benchmarks are well within the range of detectability of CTA and SKA. It should be noted that there are many limits set on decaying DM in the literature~\cite{Yuksel:2007dr,PalomaresRuiz:2007ry,Zhang:2009ut,Cirelli:2009dv,Bell:2010fk,Dugger:2010ys,Cirelli:2012ut,Murase:2012xs,Mambrini:2015sia}. These analyses give limits on the DM lifetime to be $\mathcal{O}(10^{27-28})$ s with the exception of light DM (in the MeV$-$GeV range) where the constraints are less stringent. In Ref.~\cite{Essig:2013goa}, data from HEAO-1~\cite{Gruber:1999yr}, INTEGRAL~\cite{Bouchet:2008rp}, COMPTEL~\cite{Weidenspointner:2000aq} and EGRET~\cite{Strong:2004de} were used to set limits on decaying light DM particles. Furthermore, using a microwave background analysis, one sets a limit on the lifetime of DM decaying to 30 MeV electrons and positrons~\cite{Slatyer:2016qyl} to be greater than $2.6\times 10^{25}$ s, at 95\% CL. Scanning the parameter space of the model, one 
finds many more points that can be probed by  future experiments in a wide range of photon energy. Those points are shown in the right panel of Fig.~\ref{fig4}. We note that the sharp cutoff of model points 
  in the scatter plot is due to the limits placed on the parameter space of input used in 
  generating  the  scatter plot.  
  Just as for the benchmarks of Table~\ref{tab1}, the model points in the scatter plot have enhanced leptonic (mostly tau) three body decays with an estimated flux below the limits shown in Fig.~\ref{fig2} and Fig.~\ref{fig3}.  The discovery of a monochromatic spectral line feature would be a strong evidence for a decaying dark matter since standard astrophysical processes do not produce such an effect.

\section{Conclusions}\label{sec:conc}

The search for  dark matter has resulted in both direct and indirect detection experiments. Direct detection involves either producing the DM particles directly using collider experiments such as the LHC or using large volume underground experiments such as Xenon1T and the upcoming XenonnT to measure electronic or nuclear recoil effects as DM particles scatter off xenon atoms. Most recently, Xenon1T~\cite{Aprile:2020tmw} detected some excess in electronic recoil events in the few keV region with hidden sector DM as a possible explanation (see, e.g.,~\cite{Aboubrahim:2020iwb} and the references therein).
However, a radioactive tritium background cannot be ruled out as a possible reason for such an excess. Indirect detection experiments have been reporting several signals for possible DM particles through their annihilation or decay products such as the positron excess  by the PAMELA~\cite{Adriani:2008zr} and AMS-02~\cite{Accardo:2014lma,Aguilar:2019owu,Aguilar:2019ksn} experiments. In contrast to charged particles which cannot be easily traced back to their origin due to energy losses and diffusion as those particles travel through the universe, gamma ray signals do not suffer from those uncertainties and can be a powerful tool in searching for DM. 

In this work we  have investigated the possibility of detecting a monochromatic gamma ray line resulting from the radiative  decay of a long-lived 
 MSSM/SUGRA 
neutralino $\tilde\chi^0_1$ into the hidden sector fermion $\tilde\xi^0_1$ in the process  $\tilde\chi^0_1\to\tilde\xi^0_1\gamma$.
%An analysis of this decay is given in the framework of $U(1)_X$ extended SUGRA model with two extra $U(1)$'s
%where the $U(1)_Y$ gauge field of the visible sector mixes with the hidden sector $U(1)_{X_1}$ gauge field
%via kinetic mixing and  $U(1)_{X_1}$ has kinetic mixing with $U(1)_{X_2}$. The mass growth of the hidden sector gauginos  comes about from
%the Stueckelberg mechanism and soft breaking.
% The decay 
% involves the computation of supersymmetric diagrams which include the 
% fermion-sfermion, $W$ boson-charginos, and 
% charged Higgs-charginos in the loops. 
With mixing between the visible sector
  and the hidden sector neutralino $\tilde\xi_1^0$ chosen to be ultraweak, the lifetime for the decay of the neutralino
  is much larger than the lifetime of the universe. Thus it was shown that 
  the MSSM/SUGRA neutralino of the visible sector, although unstable, can generate the full dark matter  relic density 
  of the universe. At the same time, production of  the hidden sector neutralino is suppressed 
   in the early universe because of its ultraweak interactions with the SM particles. Further, 
  the amount of hidden sector neutralino $\tilde\xi^0_1$ produced via the decay of the visible sector neutralino
  is negligible because the decay lifetime of the visible sector neutralino is significantly larger than
  the lifetime of the universe. Thus the MSSM/SUGRA neutralino of the visible sector is the dominant
  component of dark matter in the universe at current times with the hidden sector neutralino $\tilde\xi_1^0$ being a subdominant component.

To investigate in a quantitative fashion  the possibility of detection of the monochromatic 
gamma ray signal in the model, we generated a number of benchmarks given  in Table~\ref{tab1}. 
To ensure the viability of the benchmarks we investigated the diffuse gamma ray emission which can arise in this model from three body decays
such as $\tilde\chi^0_1\to\tilde\xi^0_1 f\bar f$, where the final state is dominated by $\tau^+\tau^-$.
Similarly, the DM annihilation is dominated by the channel $\tilde\chi^0_1\tilde\chi^0_1\to \tau^+\tau^-$.
These decays produce diffuse photons and we  showed that the produced photon fluxes do not exceed the current limit from Fermi-LAT allowing for  a monochromatic 
photon line to be observed in current and future experiments   with greater sensitivity 
such as SKA and CTA.  We also showed that the antiproton flux arising from these
decays and annihilation for the benchmarks makes a very suppressed contribution
to the antiproton background from astrophysical sources.
It is remarkable that despite the fact that line emissions are loop suppressed compared to diffuse emissions from three-body decays, the former can still produce a signal that may be detectable in
future experiments.

Finally, we point to the new elements of the analysis and compare our work to the existing 
literature. 
 The model discussed in section~\ref{sec:model} is an extended
 $\text{MSSM/SUGRA}\times U(1)_{X_1}\times U(1)_{X_2}$ model with 8 neutralinos 
involving kinetic mixing of three $U(1)'s$, i.e., $U(1)_Y\times U(1)_{X_1}\times U(1)_{X_2}$, 
where the visible sector communicates with the hidden sector $X_2$ only via the sector $X_1$
making the coupling of the visible sector with $X_2$ doubly suppressed. This mechanism
is helpful in allowing the visible sector neutralino to decay into the lighter neutralino  in $X_2$
with lifetime larger that the age of the universe with couplings of two $U(1)$'s of size $\mathcal{O}(10^{-12})$.
This is to be contrasted with the case of just one $U(1)_X$ which would require a mixing of 
size $\mathcal{O}(10^{-24})$. While the ultraweak mixings of size $\mathcal{O}(10^{-12})-\mathcal{O}(10^{-24})$
can be manufactured in string compactifications, we believe that mixings of size $\mathcal{O}(10^{-12})$ offer  less fine tuned solutions
than those of size $\mathcal{O}(10^{-24})$. However, this is to a degree subjective. In any case, the proposed
mechanism of section~\ref{sec:model} is new and offers another solution to achieving ultraweak mixings. 
Regarding sections~\ref{sec:loop} and~\ref{sec:loopwidth}, here the first complete analysis of $W$-charginos, charged Higgs-charginos,
and fermion-sfermions exchange contributions in the loops is given. The parameter space of the
model is constrained by the astrophysical and laboratory constraints which include the relic density, the gamma ray flux and antiproton flux. The monochromatic signal is analysed in detail. 
%Also discussed are the prospects of detection of dark matter in direct detection experiments. 

We now compare our analysis to previous works. 
While decaying dark matter has been discussed by various authors there is currently no 
focused work on the decaying neutralino in the framework we discuss. Thus, while the work of~\cite{Sierra:2009zq} 
 discusses decaying neutralino as dark matter, the analysis is within the  framework of $R$-parity violation and has no overlap with our work. In our analysis, $R$-parity is strictly conserved and the decay of the neutralino arises at the loop level.  The work of Ref.~\cite{Ibarra:2009bm} does consider mSUGRA model points but is restricted to leptonic three-body decays to explain the positron excess and does not discuss the
 radiative two body decay of the neutralino. 
 Ref.~\cite{Garny:2010eg} discusses radiative decay but the analysis is in a non-supersymmetric framework with a 
 brief discussion of the SUSY case. 
Further,  the model assumes the dark matter to be leptophilic to avoid hadronic decays of DM.  
In our analysis a natural suppression of hadronic channels occurs due to a light stau as discussed in the text.  Moreover, our analysis is consistent with the observed Higgs boson mass and 
the relic density constraint which are overlooked in some of the previous analyses. Finally, the analyses of Figs.~\ref{fig3} and~\ref{fig4} takes account of the most recent experimental limits from Fermi-LAT and CTA and discusses the prospects of the new SKA experiment.

\noindent
{\bf Acknowledgments:} The research of AA and MK was supported by the BMBF under contract 05H18PMCC1, while the research of PN was supported in part by the NSF Grant PHY-1913328.

%\begin{appendices}
\appendix

\section*{\Huge Appendices}

\section{Interactions that enter in the analysis of the neutralino radiative decay}\label{sec:loop}

The radiative neutralino decay $\tilde\chi_1^0\to \tilde\xi_1^0 \gamma$ is a loop suppressed process and the loop diagrams contributing to it are given in Fig.~\ref{fig1}. 
The loop diagrams involve the following interactions: $WW\gamma$, chargino-chargino-$\gamma$, neutralino-chargino-$W$, neutralino-fermion-sfermion and neutralino-chargino-charged Higgs. 
We discuss them below. 

The $WW\gamma$ coupling is obtained from the triboson Lagrangian
\begin{align}
\mathcal{L}_{V} &= 
i e \cot{\theta_W}
\left[
\left(\partial^\mu W^\nu -\partial^\nu W^\mu\right)
 W^\dagger_\mu Z_\nu -
\left(\partial^\mu W^{\nu\dagger} -\partial^\nu W^{\mu\dagger}\right)
 W_\mu Z_\nu +
W_\mu W^\dagger_\nu\left(\partial^\mu Z^\nu -\partial^\nu Z^\mu\right)
\right]\nonumber\\
&+i e \left[
\left(\partial^\mu W^\nu -\partial^\nu W^\mu\right)
 W^\dagger_\mu A_\nu -
\left(\partial^\mu W^{\nu\dagger} -\partial^\nu W^{\mu\dagger}\right)W_\mu A_\nu + W_\mu W^\dagger_\nu\left(\partial^\mu A^\nu -\partial^\nu A^\mu\right)
\right].
\end{align}
The chargino-chargino-$\gamma$ coupling is given by 
\begin{align}
\mathcal{L}_{\tilde\chi\gamma}= -e A_\mu  \bar \chi^C \gamma^\mu \chi^C\,,
\end{align}
while the neutralino-chargino-$W$ coupling is given by 
 \begin{align}
\mathcal{L}_{\tilde\chi^0\tilde\chi W}= g_2 W_\mu^{-} \bar{\chi_i}^0 \gamma^\mu \left( O^L_{ij} P_L + O^R_{ij} P_R\right) \chi_j^C + \text{h.c.}, 
\label{h-chargino.1}
\end{align}
with
\begin{align}
O^L_{ij}&= -\frac{1}{\sqrt 2}X_{i4} V^*_{j2} + X_{i2} V^*_{j1},\nonumber\\
O^R_{ij}&= \frac{1}{\sqrt 2}X^*_{i3} U_{j2} + X^*_{i2} U_{j1},
\label{O-eq}
\end{align}
where $X$ diagonalizes the neutralino mass matrix $M_N$ of Eq.~(\ref{nmatrix1}) so that 
\begin{align}
X^* M_{N} X^{-1}=\text{ diag} ( m_{\tilde\chi_1^0},
m_{\tilde\chi_2^0},m_{\tilde\chi_3^0},m_{\tilde\chi_4^0}, m_{\tilde\chi_5^0}, m_{\tilde\chi_6^0}, m_{\tilde\chi_7^0}, m_{\tilde\chi_8^0}),
\end{align}
and $\tilde\xi^0_4\equiv\tilde\chi_5^0,~\tilde\xi^0_3\equiv\tilde\chi_6^0,~\tilde\xi^0_2\equiv\tilde\chi_7^0,~\tilde\xi^0_1\equiv\tilde\chi_8^0$.
The Majorana phases are absorbed into $X$ so that the masses are non-negative. The $U$ and $V$ are the matrices that diagonalize the chargino mass matrix so that 
\begin{align}
 U^* M_C V^{-1}= \text{ diag} (m_{\tilde\chi_1^{\pm}},
m_{\tilde\chi_2^{\pm}}),
\end{align}
where $M_C$ is the chargino mass matrix. 
 
\subsection{The effective Lagrangian for $\bar q \tilde q_i \tilde\chi_j^0$ interaction} 

The quark-squark-neutralino, $\bar q \tilde q_i \chi_j^0$, interaction at the tree level  is given by   
\begin{equation}
\mathcal{L}_{q\tilde q\tilde\chi^0}= g\bar d (K^d_{ij} P_R + M^d_{ij} P_L)\tilde\chi_j^0 \tilde d_i
+g \bar u (K^u_{ij} P_R + M^u_{ij} P_L)\tilde\chi_j^0\tilde u_i + \text{h.c.}
\label{neutralino-lag}
\end{equation}
Here $i=1,2$ stand for the two squark states (first generation sup and sdown), $j=1\cdots 8$ stand for the 8 neutralino states  and where the couplings in Eq.~(\ref{neutralino-lag}) are given by
\begin{equation}
\begin{aligned}
K^d_{ij}&= -\sqrt 2 (\beta_{dj}D_{d1i} + \alpha_{dj}^* D_{d2i}) , \\
K^u_{ij}&= -\sqrt 2 (\beta_{uj}D_{u1i} + \alpha_{uj}^* D_{u2i}), \\
M^d_{ij}&= -\sqrt 2 (\alpha_{dj}D_{d1i} - \gamma_{dj} D_{d2i}),  \\
M^u_{ij}&= -\sqrt 2 (\alpha_{uj}D_{u1i} - \gamma_{uj} D_{u2i}).  
\end{aligned}
\label{KM}
\end{equation}
The quantity $D_q$ (with $q=u,d$) in Eq.~(\ref{KM}) is the matrix that diagonalizes the squark mass square matrix, $M_{\tilde{q}}^2$, i.e.,
\begin{align}
D_{q}^\dagger M_{\tilde{q}}^2 D_q={\rm diag}(m_{\tilde{q}1}^2,
              m_{\tilde{q}2}^2),
\end{align} 
where $M_{\tilde{q}}^2$ is given by
\begin{align}
\left(\begin{matrix} 
M_{\tilde{Q}}^2+m{_q}^2+M_{Z}^2(\frac{1}{2}-Q_q
\sin^2\theta_W)\cos2\beta & m_q(A_{q}-\mu R_q) \cr
   	          	m_q(A_{q}-\mu R_q) & M_{\tilde{U}}^2+m{_q}^2+M_{Z}^2 Q_q \sin^2\theta_W \cos2\beta
\end{matrix}\right).
\end{align}
Here $Q_q=2/3(-1/3)$ and $R_q=\cot\beta(\tan\beta)$ for $q=u(d)$, and $m_q$ is the quark mass. The various other quantities appearing in Eq.~(\ref{KM}) are defined as follows~\cite{Ibrahim:2004gb}
\begin{equation}
\begin{aligned}
\alpha_{d(u)k}& =\frac{g m_{d(u)}X_{3(4)k}}{2m_W\cos\beta(\sin\beta)},~~~\beta_{d(u)k}=eQ_{d(u)}X_{1k}^{'*} +\frac{g}{\cos\theta_W} X_{2k}^{'*}
(T_{3d(u)}-Q_{d(u)}\sin^2\theta_W), \\
\gamma_{d(u)k}&=eQ_{d(u)} X_{1k}'-\frac{gQ_{d(u)}\sin^2\theta_W}{\cos\theta_W}X_{2k}',
\end{aligned}
\end{equation}
where $T_{3d} =-\frac{1}{2}$, $T_{3u}=\frac{1}{2}$, $Q_d= -\frac{1}{3}$, $Q_u= \frac{2}{3}$ and the $X'$'s are given in terms of $X$, the matrix that diagonalizes the neutralino mass matrix, as
\begin{equation}
\begin{aligned}
X'_{1k}&=X_{1k}\cos\theta_W +X_{2k}\sin\theta_W, \\
X'_{2k}&=-X_{1k}\sin\theta_W +X_{2k}\cos\theta_W.
\end{aligned}
\end{equation}
The same analysis can be easily generalized to include the second and third generation quarks and squarks. 
To extract the interactions of the MSSM neutralino $\tilde\chi^0_1$ from Eq.~(\ref{neutralino-lag}) we set $j=1$ while setting $j=8$ allows one to obtain the interactions of the LSP, $\tilde\xi^0_1$. 

\subsection{The effective Lagrangian for $\bar\ell \tilde\chi_i^0\tilde\ell_j$ interaction}

To determine the elements of this interaction, we start by writing the slepton mass square matrix. It is given by   
\begin{equation}
M_{\tilde{{\ell}}}^2=\left(\begin{matrix} M_{\tilde{L}}^2+m{_{{\ell}}}^2
-M_{Z}^2(\frac{1}{2}-
\sin^2\theta_W)\cos2\beta & m_{{\ell}}(A_{{\ell}}-\mu \tan\beta) \cr
   	          	m_{{\ell}}(A_{{\ell}}-{\mu} \tan\beta) & 
		M_{\tilde R}^2+m{_{\ell}}^2-M_{Z}^2  \sin^2\theta_W \cos2\beta 
\end{matrix}\right),
\end{equation}
which is a hermitian matrix and can be diagonalized by the unitary transformation
\begin{equation}
D_{{\ell}}^\dagger M_{\tilde{{\ell}}}^2 D_{{\ell}}={\rm diag}(m_{\tilde{{\ell}}1}^2,
              m_{\tilde{{\ell}}2}^2).
\end{equation}
The neutralino-lepton-slepton  interaction in the mass
diagonal basis is defined by 
\begin{equation}
-\mathcal{L}_{\ell\tilde\ell\tilde\chi^0}=\bar\ell(F_{ij}P_L+Z_{ij}P_R)\tilde\chi^0_i\tilde\ell_j + \text{h.c.},
\label{slepton-lepton}
\end{equation}
where the couplings $F_{ij}$ and $Z_{ij}$ are given by
\begin{equation}
\begin{aligned}
F_{ij}&=\sqrt{2}(\alpha_{\ell i}D_{\ell 1j}-\gamma_{\ell i}D_{\ell 2j}), \\
Z_{ij}&=\sqrt{2}(\beta_{\ell i}D_{\ell 1j}-\delta_{\ell i}D_{\ell 2j}),
\end{aligned}
\label{last}
\end{equation}
and the quantities $\alpha$, $\beta$, $\gamma$ and $\delta$ are
\begin{equation}
\begin{aligned}
\alpha_{{\ell} i}&=\frac{gm_{{\ell}}X_{3i}}{2m_W  \cos{\beta}}, ~~~\beta_{{\ell} i}=eQ_{{\ell}}X_{1i}^{'*} +\frac{g}{\cos{\theta_W}}
 X_{2i}^{'*}(T_{3{\ell}}-Q_{{\ell}}\sin^{2}{\theta_W}), \\
\gamma_{{\ell} i}&=eQ_{{\ell}}X_{1i}^{'} -\frac{gQ_{{\ell}}\sin^{2}{\theta_W}}
{\cos{\theta_W}}  X_{2i}^{'}, ~~~\delta_{{\ell} i}=-\frac{gm_{{\ell}}X_{3i}^*}{2m_W  \cos{\beta}}.
\end{aligned}
\end{equation}
Again, the $\tilde\chi_1^0\ell\tilde \ell$ interaction can be gotten by setting $i=1$ and $\tilde\xi_1^0\ell\tilde \ell$ is obtained for $i=8$.

\subsection{The effective Lagrangian for $H^-\tilde\chi^0_i\tilde\chi^+_j$ interaction}

The effective Lagrangian that describes the charged Higgs-chargino-neutralino vertex is given by
  \begin{equation}
{\cal {L}}_{H\tilde\chi^0\tilde\chi^+}=H^-\bar{\tilde\chi}^0_i(\alpha^S_{ij}+\gamma_{5}\alpha^P_{ij})\tilde\chi^+_j+ \text{h.c.},
\label{h-charge-neu.1}
\end{equation}
where $i=1\cdots 8$, $j=1,2$ and 
\begin{equation}
\begin{aligned}
\alpha^{S}_{ij}&=\frac{1}{2}\xi'_{ij}\sin{\beta}+
\frac{1}{2}\xi_{ij}\cos{\beta}, \\
\alpha^{P}_{ij}&=\frac{1}{2}\xi'_{ij}\sin{\beta}
-\frac{1}{2} \xi_{ij} \cos{\beta}.
\label{spcoupling}
\end{aligned}
\end{equation}
 In the above, 
  $\xi_{ij}$ and $\xi_{ij}'$ are given by 
\begin{equation}
\begin{aligned}
\xi_{ij}&=-gX_{4i}V_{j1}^* 
-\frac{g}{\sqrt 2} X_{2i} V_{j2}^* 
-\frac{g}{\sqrt 2} \tan\theta_W X_{1i}V_{j2}^*, \\
\xi_{ij}'&=-gX_{3i}^*U_{j1} 
+\frac{g}{\sqrt 2} X_{2i}^* U_{j2} 
+\frac{g}{\sqrt 2} \tan\theta_W X_{1i}^*U_{j2}.
\end{aligned}
\end{equation}
To obtain the interactions of $\tilde\chi^0_1$ we set $i=1$ and for $\tilde\xi^0_1$ we set $i=8$.

\section{The flux from relevant astrophysical processes}\label{sec:id}

The main source of gamma ray photons reaching terrestrial
detectors arises from the galactic center (GC). Those photons can be prompt, i.e. the result of internal bremsstrahlung, or come from the decay of SM hadrons or decay/annihilation of DM particles. Secondary sources of photons are due to inverse Compton scattering (ICS), bremsstrahlung and  synchrotron emission. In this work, prompt emissions and ICS dominate for the energy range and observational region of interest. The photon flux from the GC depends on the DM density profile which we consider to be of the Navarro, Frenk and White (NFW) form~\cite{Navarro:1995iw}. In this Appendix we give details of the calculations of the gamma ray flux from prompt photons, ICS and IGRB as well as the antiproton flux. 

\subsection{Prompt $\gamma$ ray photons}

SM final states (leptons, quarks, $W^{\pm}$, $Z$) resulting from the decay or annihilation of DM particles in the GC will give rise to a continuum and diffuse photon spectrum. Photons are produced either by final state radiation off charged particles or by the decay of hadrons, such as pions. The spectrum has a kinematic endpoint at half the DM mass for decay processes and at the dark matter mass for annihilation. \\
The differential photon flux for an observation region $\Delta\Omega$ is given by
\begin{equation}
\frac{d\Phi}{dE\Delta\Omega}(E_{\gamma})=\frac{r_{\odot}}{4\pi}\frac{\rho_{\odot}}{m_{\tilde\chi^0_1}}\bar{J}\sum_i \frac{1}{\tau_i}\frac{dN_i}{dE_{\gamma}},
\label{sdecay}
\end{equation} 
where $dN_i/dE_{\gamma}$ is the photon spectrum produced from the decay into some final state $i$ and is determined using \code{PYTHIA}~\cite{Sjostrand:2007gs}\footnote{{QCD uncertainties in the modeling of the gamma ray spectrum in DM annihilation from event generators have been analyzed in Ref.~\cite{Amoroso:2018qga}. Depending on the DM mass and the photon energy-to DM mass ratio, the uncertainties can range from $\sim 30\%$ to $\sim 40\%$. Taking those uncertainties into consideration does not change our numerical analysis in any drastic way.}}, $\tau_i$ is the DM lifetime, $\bar{J}$ is the average $J$ factor for prompt photons given by
\begin{equation}
\bar{J}=\frac{4}{\Delta\Omega}\iint db\,d\ell \cos b\,J[\theta(b,\ell)],
\end{equation} 
in the $b\times\ell$ region with $(b,\ell)$ being the galactic coordinates (latitude $b$ and longitude $\ell$) and
\begin{equation}
\Delta\Omega=4\iint db \,d\ell\cos b.
\end{equation}
The $J$ factor, $J(\theta)$ is calculated with \code{PPPC4DMID}~\cite{Ciafaloni:2010ti,Cirelli:2010xx} where $\cos\theta= \cos b\cos\ell$ near the galactic plane. Note that $\theta$ is the aperture angle between the direction of the line of sight and the axis connecting
the Earth to the GC. $r_{\odot}$ is the location of the Sun in the galactic plane which we take to be 8.5 kpc and $\rho_{\odot}=0.4$ GeV/cm$^3$ is the DM mass density at $r_{\odot}$.  

One can write a similar expression for the photon flux  for annihilation processes
\begin{equation}
\frac{d\Phi}{dE\Delta\Omega}(E_{\gamma})=\frac{r_{\odot}}{4\pi}\frac{1}{2}\left(\frac{\rho_{\odot}}{m_{\tilde\chi^0_1}}\right)^2\bar{J}\sum_i \langle\sigma v\rangle_i\frac{dN_i}{dE_{\gamma}},
\label{sann}
\end{equation}  
where $\langle\sigma v\rangle_i$ is the DM thermally averaged cross-section into some final states $i$. This diffuse and broad spectrum of photon emission from DM particles is hard to disentangle from astrophysical processes which constitute an overwhelming background. 

The direct decay/annihilation of DM particles to photons will produce a sharp spectral line which can be easily detected over the otherwise smooth background from astrophysical emissions. Such a spectral line would be centered at
\begin{equation}
E_{\gamma}=\frac{m_{\tilde\chi^0_1}}{2}\left(1-\frac{m^2_{\tilde\xi^0_1}}{m^2_{\tilde\chi^0_1}}\right),
\label{egdecay}
\end{equation} 
for the decay $\tilde\chi^0_1\to\tilde\xi^0_1\gamma$ and at
$E_\gamma= m_{\tilde\chi^0_1}$ 
for annihilation $\tilde\chi^0_1\tilde\chi^0_1\to\gamma\gamma$. In this case, the photon spectrum becomes $dN/dE_{\gamma}=\delta(E-E_{\gamma})$ for decay and $2\delta(E-E_{\gamma})$ for annihilation. 

\subsection{Inverse Compton scattering}

Photon flux signals generated in the GC are generally the most intense but are plagued with high levels of astrophysical background processes. At higher latitudes the signal becomes weaker and so does the contamination from the background. Energetic SM final states such as electrons arising from the decay or annihilation of DM particles can scatter off the interstellar light and the cosmic microwave background (CMB) sending those soft photons into the gamma ray energy range. Such a process is known as ICS. The photon signal from ICS comes from every direction in the diffusion volume of the galactic halo including high latitudes where astrophysical processes are subdominant and thus suffers from less uncertainties compared to the signal from the GC. 

Products of DM decay or annihilation can be grouped into primary, $p$, and secondary, $s$. Primary particles are the result of the immediate decay/annihilation while secondary particles are stable SM particles resulting from the cascade decays of the primary particles. For ICS, electrons and positrons are the relevant secondary particles responsible for this scattering process. The differential ICS photon flux in the case of DM decay is given by
\begin{equation}
\frac{d\Phi_{\rm ICS}}{dE_{\gamma}\Delta\Omega}=\frac{r_{\odot}}{4\pi}\frac{\rho_{\odot}}{m_{\tilde\chi^0_1}}\frac{1}{E^2_{\gamma}}\int_{m_s}^{m_{\tilde\chi^0_1}/2}dE_s\sum_i \frac{1}{\tau_i}\frac{dN^i_{s}}{dE}(E_s)F_{\rm ICS}(E_{\gamma},E_s,b,\ell),
\label{icsdecay}
\end{equation}  
where $E_s$ and $m_s$ are the injection energy and mass of the secondary particle (in this case $e^{\pm}$) and $F_{\rm ICS}$ is a Halo function determined using the NFW profile and a set of propagation parameters~\cite{Buch:2015iya}. For the DM annihilation case, the differential flux takes the form
\begin{equation}
\frac{d\Phi_{\rm ICS}}{dE_{\gamma}\Delta\Omega}=\frac{r_{\odot}}{4\pi}\left(\frac{\rho_{\odot}}{m_{\tilde\chi^0_1}}\right)^2\frac{1}{2E^2_{\gamma}}\int_{m_s}^{m_{\tilde\chi^0_1}}dE_s\sum_i \langle\sigma v\rangle_i\frac{dN^i_{s}}{dE}(E_s)F_{\rm ICS}(E_{\gamma},E_s,b,\ell).
\label{icsann}
\end{equation}

\subsection{Antiproton flux}\label{sec:antipflux}

The production and propagation of cosmic rays in the galaxy is a complicated process since it requires a modeling of diffusion, energy loss and annihilation effects as the charged particles navigate through the magnetic field of the galaxy. This process is generally described by a transport equation which arises from a stationary two-zone diffusion model with cylindrical boundary conditions~\cite{Ginzburg:1990sk,Ibarra:2009bm}. For the case of antiprotons of mass $m_p$ resulting from DM decay, the solution to the transport equation at the position of the solar system is given by
\begin{equation}
f(E)=\frac{1}{\tau m_{\tilde\chi^0_1}}\int_0^{m_{\tilde\chi^0_1}-m_p}dE' \,G(E,E') \frac{dN(E')}{dE'},
\end{equation} 
where the Green's function $G(E,E')$ contains all the astrophysical information. The antiproton differential flux for the case of DM decay can be expressed as~\cite{Boudaud:2014qra}
\begin{equation}
\frac{d\Phi^{\rm IS}_p}{dE'}=\frac{v_p}{4\pi}\frac{\rho_{\odot}}{m_{\tilde\chi^0_1}}G(E,E')\sum_i\frac{1}{\tau_i}\frac{dN_i}{dE'},
\label{antipd}
\end{equation}
and for the annihilation case
\begin{equation}
\frac{d\Phi^{\rm IS}_p}{dE'}=\frac{v_p}{4\pi}\frac{1}{2}\left(\frac{\rho_{\odot}}{m_{\tilde\chi^0_1}}\right)^2 G(E,E')\sum_i\langle\sigma v\rangle_i\frac{dN_i}{dE'}.
\label{antipa}
\end{equation}
The superscript `IS' in Eqs.~(\ref{antipd}) and~(\ref{antipa}) means that these expressions are for the interstellar flux. The measurement of the antiproton flux is done at the top of the atmosphere (TOA) and such a value is affected by solar modulations. The solar wind decreases the kinetic energy and momentum of antiprotons and so in order to compare our results with AMS-02 experiment~\cite{Aguilar:2016kjl}, we determine the TOA flux as
\begin{equation}
\Phi_p^{\rm TOA}(E_{\rm TOA})=\left(\frac{2m_p E_{\rm TOA}+E^2_{\rm TOA}}{2m_p E_{\rm IS}+E^2_{\rm IS}}\right)\Phi^{\rm IS}_p(E_{\rm IS}),
\label{mod}
\end{equation}
where the IS and TOA kinetic energies are related by $E_{\rm IS}=E_{\rm TOA}+\phi_M$, with the modulation parameter $\phi_M$  taken to be 500 MV which corresponds to minimum solar activity.

\subsection{The isotropic gamma ray background}\label{sec:igrb}

As mentioned earlier, measurements of the isotropic gamma ray background (IGRB) produce stringent limits on the DM decay lifetime~\cite{Abdo:2010nc,Ackermann:2015lka} and so it is important to consider this component and make sure that the photon flux from DM decay and annihilation is within the current experimental limits. 
The IGRB is the faint component of the photon flux measured at high galactic latitudes ($|b|>20^{\circ}$) with main contributions from active galactic nuclei and star-forming galaxies~\cite{Ackermann:2014usa}. Together with extragalactic contributions due to energetic cosmic rays generating electromagnetic cascades or interactions with the galactic gas, the measured IGRB spectrum leaves very little room for exotic components such as DM decay or annihilation~\cite{Blanco:2018esa}. The flux due to prompt photons and ICS from DM decay or annihilation can be calculated for $|b|>20^{\circ}$ using Eqs.~(\ref{sdecay}),~(\ref{sann}),~(\ref{icsdecay}) and~(\ref{icsann}). Next, we discuss calculating the photon flux of extragalactic (EG) origin.

The differential prompt photon flux of extragalactic gamma rays due to DM decay measured at a redshift $z$ is given by
\begin{align}
\frac{d\Phi^{\rm EG}_{\gamma}}{dE_{\gamma}}(E_{\gamma},z)=c\int_0^{\infty} dz'\frac{(1+z)^2}{H(z')(1+z')^3}\frac{\rho_{\chi}(z')}{m_{\tilde\chi}}e^{-\tau(E_{\gamma},z,z')}\sum_i\frac{1}{\tau_i}\frac{dN^i}{dE_{\gamma}}(E'_{\gamma}),
\end{align}
where $H(z')\equiv H_0\sqrt{\Omega_{\rm m}(1+z')^3+(1-\Omega_{\Lambda})}$, $\rho_\chi(z')=\rho_0(1+z')^3$ and $E_{\gamma}'\equiv E_{\gamma}(1+z')/(1+z)$, where the latter relates the photon energy $E_{\gamma}$ at $z$ with the photon energy $E_{\gamma}'$ at $z'$. Here, $\Omega_{\rm m}=0.31$, $\Omega_{\Lambda}=0.69$, $H_0=67.7$ km/s and $\rho_0=1.15\times 10^{-6}$ GeV/cm$^3$. The quantity $\tau(E_{\gamma},z,z')$ is the optical depth of the universe between the redshifts $z$ and $z'$. This takes into account the absorption effects due to pair production.  Since the measurement of the flux is done at $z=0$, the differential flux for decay is given by
\begin{align}
\frac{d\Phi^{\rm EG}_{\gamma}}{dE_{\gamma}}(E_{\gamma})=\frac{c}{m_{\tilde\chi}}\int_0^{\infty} dz'\frac{\rho_{\chi}(z')}{H(z')(1+z')^3}e^{-\tau(E_{\gamma},z')}\sum_i\frac{1}{\tau_i}\frac{dN^i}{dE_{\gamma}}(E'_{\gamma}),
\end{align}
and for the annihilation case
\begin{align}
\frac{d\Phi^{\rm EG}_{\gamma}}{dE_{\gamma}}(E_{\gamma})=\frac{c}{2m^2_{\tilde\chi}}\int_0^{\infty} dz'\frac{\rho_{\chi}(z')^2}{H(z')(1+z')^3}B(z')e^{-\tau(E_{\gamma},z')}\sum_i\langle\sigma v\rangle_i\frac{dN^i}{dE_{\gamma}}(E'_{\gamma}),
\end{align}
where $B(z')$ is a cosmological boost factor which corrects for DM clustering and is only relevant for the annihilation case. 

The second contribution to the IGRB is from ICS processes on CMB photons and is more complicated. The calculations require knowledge of the electron-positron number density at redshift $z'$ as well as the background bath photons (the CMB being the dominant component). We follow the recipe outlined in Ref.~\cite{Cirelli:2010xx} to calculate this contribution.


\begin{thebibliography}{99}

%\cite{Adriani:2008zr}
\bibitem{Adriani:2008zr}
O.~Adriani \textit{et al.} [PAMELA],
%``An anomalous positron abundance in cosmic rays with energies 1.5-100 GeV,''
Nature \textbf{458}, 607-609 (2009)
doi:10.1038/nature07942
[arXiv:0810.4995 [astro-ph]].
%2303 citations counted in INSPIRE as of 11 Dec 2020

%\cite{Aguilar:2007yf}
\bibitem{Aguilar:2007yf}
M.~Aguilar \textit{et al.} [AMS 01],
%``Cosmic-ray positron fraction measurement from 1 to 30-GeV with AMS-01,''
Phys. Lett. B \textbf{646}, 145-154 (2007)
doi:10.1016/j.physletb.2007.01.024
[arXiv:astro-ph/0703154 [astro-ph]].
%352 citations counted in INSPIRE as of 10 Dec 2020

%\cite{FermiLAT:2011ab}
\bibitem{FermiLAT:2011ab}
M.~Ackermann \textit{et al.} [Fermi-LAT],
%``Measurement of separate cosmic-ray electron and positron spectra with the Fermi Large Area Telescope,''
Phys. Rev. Lett. \textbf{108}, 011103 (2012)
doi:10.1103/PhysRevLett.108.011103
[arXiv:1109.0521 [astro-ph.HE]].
%668 citations counted in INSPIRE as of 07 Dec 2020

%\cite{Accardo:2014lma}
\bibitem{Accardo:2014lma}
L.~Accardo \textit{et al.} [AMS],
%``High Statistics Measurement of the Positron Fraction in Primary Cosmic Rays of 0.5\textendash{}500 GeV with the Alpha Magnetic Spectrometer on the International Space Station,''
Phys. Rev. Lett. \textbf{113}, 121101 (2014)
doi:10.1103/PhysRevLett.113.121101
%558 citations counted in INSPIRE as of 10 Dec 2020

%\cite{Aguilar:2019owu}
\bibitem{Aguilar:2019owu}
M.~Aguilar \textit{et al.} [AMS],
%``Towards Understanding the Origin of Cosmic-Ray Positrons,''
Phys. Rev. Lett. \textbf{122}, no.4, 041102 (2019)
doi:10.1103/PhysRevLett.122.041102
%90 citations counted in INSPIRE as of 27 Nov 2020

%\cite{Aguilar:2019ksn}
\bibitem{Aguilar:2019ksn}
M.~Aguilar \textit{et al.} [AMS],
%``Towards Understanding the Origin of Cosmic-Ray Electrons,''
Phys. Rev. Lett. \textbf{122}, no.10, 101101 (2019)
doi:10.1103/PhysRevLett.122.101101
%47 citations counted in INSPIRE as of 27 Nov 2020

%\cite{Aguilar:2016kjl}
\bibitem{Aguilar:2016kjl}
M.~Aguilar \textit{et al.} [AMS],
%``Antiproton Flux, Antiproton-to-Proton Flux Ratio, and Properties of Elementary Particle Fluxes in Primary Cosmic Rays Measured with the Alpha Magnetic Spectrometer on the International Space Station,''
Phys. Rev. Lett. \textbf{117}, no.9, 091103 (2016)
doi:10.1103/PhysRevLett.117.091103
%357 citations counted in INSPIRE as of 11 Dec 2020

%\cite{Ibarra:2013cra}
\bibitem{Ibarra:2013cra}
A.~Ibarra, D.~Tran and C.~Weniger,
%``Indirect Searches for Decaying Dark Matter,''
Int. J. Mod. Phys. A \textbf{28}, 1330040 (2013)
doi:10.1142/S0217751X13300408
[arXiv:1307.6434 [hep-ph]].
%112 citations counted in INSPIRE as of 05 Dec 2020

%\cite{Bull:2018lat}
\bibitem{Bull:2018lat}
A.~Weltman, P.~Bull, S.~Camera, K.~Kelley, H.~Padmanabhan, J.~Pritchard, A.~Raccanelli, S.~Riemer-S\o{}rensen, L.~Shao and S.~Andrianomena, \textit{et al.}
%``Fundamental physics with the Square Kilometre Array,''
Publ. Astron. Soc. Austral. \textbf{37}, e002 (2020)
doi:10.1017/pasa.2019.42
[arXiv:1810.02680 [astro-ph.CO]].
%73 citations counted in INSPIRE as of 18 Dec 2020

%\cite{Ghosh:2020ipv}
\bibitem{Ghosh:2020ipv}
A.~Ghosh, A.~Kar and B.~Mukhopadhyaya,
%``Search for decaying heavy dark matter in an effective interaction framework: a comparison of $\gamma$-ray and radio observations,''
JCAP \textbf{09}, 003 (2020)
doi:10.1088/1475-7516/2020/09/003
[arXiv:2001.08235 [hep-ph]].
%1 citations counted in INSPIRE as of 18 Dec 2020

%\cite{Abdalla:2020gea}
\bibitem{Abdalla:2020gea}
H.~Abdalla \textit{et al.} [Cherenkov Telescope Array Consortium],
%``Sensitivity of the Cherenkov Telescope Array for probing cosmology and fundamental physics with gamma-ray propagation,''
[arXiv:2010.01349 [astro-ph.HE]].
%4 citations counted in INSPIRE as of 11 Dec 2020

%\cite{Aharonian:2008aa}
\bibitem{Aharonian:2008aa}
F.~Aharonian \textit{et al.} [H.E.S.S.],
%``The energy spectrum of cosmic-ray electrons at TeV energies,''
Phys. Rev. Lett. \textbf{101}, 261104 (2008)
doi:10.1103/PhysRevLett.101.261104
[arXiv:0811.3894 [astro-ph]].
%643 citations counted in INSPIRE as of 01 Dec 2020

%\cite{Aharonian:2009ah}
\bibitem{Aharonian:2009ah}
F.~Aharonian \textit{et al.} [H.E.S.S.],
%``Probing the ATIC peak in the cosmic-ray electron spectrum with H.E.S.S,''
Astron. Astrophys. \textbf{508}, 561 (2009)
doi:10.1051/0004-6361/200913323
[arXiv:0905.0105 [astro-ph.HE]].
%602 citations counted in INSPIRE as of 01 Dec 2020

%\cite{Ahnen:2017pqx}
\bibitem{Ahnen:2017pqx}
M.~L.~Ahnen \textit{et al.} [MAGIC],
%``Indirect dark matter searches in the dwarf satellite galaxy Ursa Major II with the MAGIC Telescopes,''
JCAP \textbf{03}, 009 (2018)
doi:10.1088/1475-7516/2018/03/009
[arXiv:1712.03095 [astro-ph.HE]].
%18 citations counted in INSPIRE as of 12 Dec 2020

%\cite{Ahnen:2016qkx}
\bibitem{Ahnen:2016qkx}
M.~L.~Ahnen \textit{et al.} [MAGIC and Fermi-LAT],
%``Limits to Dark Matter Annihilation Cross-Section from a Combined Analysis of MAGIC and Fermi-LAT Observations of Dwarf Satellite Galaxies,''
JCAP \textbf{02}, 039 (2016)
doi:10.1088/1475-7516/2016/02/039
[arXiv:1601.06590 [astro-ph.HE]].
%277 citations counted in INSPIRE as of 27 Nov 2020

%\cite{Archambault:2017wyh}
\bibitem{Archambault:2017wyh}
S.~Archambault \textit{et al.} [VERITAS],
%``Dark Matter Constraints from a Joint Analysis of Dwarf Spheroidal Galaxy Observations with VERITAS,''
Phys. Rev. D \textbf{95}, no.8, 082001 (2017)
doi:10.1103/PhysRevD.95.082001
[arXiv:1703.04937 [astro-ph.HE]].
%68 citations counted in INSPIRE as of 27 Nov 2020

%\cite{Zitzer:2015eqa}
\bibitem{Zitzer:2015eqa}
B.~Zitzer [VERITAS],
%``A Search for Dark Matter from Dwarf Galaxies using VERITAS,''
PoS \textbf{ICRC2015}, 1225 (2016)
doi:10.22323/1.236.1225
[arXiv:1509.01105 [astro-ph.HE]].
%32 citations counted in INSPIRE as of 27 Nov 2020

%\cite{Garny:2010eg}
\bibitem{Garny:2010eg}
M.~Garny, A.~Ibarra, D.~Tran and C.~Weniger,
%``Gamma-Ray Lines from Radiative Dark Matter Decay,''
JCAP \textbf{01}, 032 (2011)
doi:10.1088/1475-7516/2011/01/032
[arXiv:1011.3786 [hep-ph]].
%40 citations counted in INSPIRE as of 27 Nov 2020

%\cite{Consortium:2010bc}
\bibitem{Consortium:2010bc}
M.~Actis \textit{et al.} [CTA Consortium],
%``Design concepts for the Cherenkov Telescope Array CTA: An advanced facility for ground-based high-energy gamma-ray astronomy,''
Exper. Astron. \textbf{32}, 193-316 (2011)
doi:10.1007/s10686-011-9247-0
[arXiv:1008.3703 [astro-ph.IM]].
%835 citations counted in INSPIRE as of 11 Dec 2020

%\cite{Abdo:2010nz}
\bibitem{Abdo:2010nz}
A.~A.~Abdo \textit{et al.} [Fermi-LAT],
%``The Spectrum of the Isotropic Diffuse Gamma-Ray Emission Derived From First-Year Fermi Large Area Telescope Data,''
Phys. Rev. Lett. \textbf{104}, 101101 (2010)
doi:10.1103/PhysRevLett.104.101101
[arXiv:1002.3603 [astro-ph.HE]].
%555 citations counted in INSPIRE as of 07 Dec 2020

%\cite{Yuksel:2007dr}
\bibitem{Yuksel:2007dr}
H.~Yuksel and M.~D.~Kistler,
%``Circumscribing late dark matter decays model independently,''
Phys. Rev. D \textbf{78}, 023502 (2008)
doi:10.1103/PhysRevD.78.023502
[arXiv:0711.2906 [astro-ph]].
%82 citations counted in INSPIRE as of 01 Dec 2020

%\cite{PalomaresRuiz:2007ry}
\bibitem{PalomaresRuiz:2007ry}
S.~Palomares-Ruiz,
%``Model-Independent Bound on the Dark Matter Lifetime,''
Phys. Lett. B \textbf{665}, 50-53 (2008)
doi:10.1016/j.physletb.2008.05.040
[arXiv:0712.1937 [astro-ph]].
%82 citations counted in INSPIRE as of 11 Dec 2020

%\cite{Zhang:2009ut}
\bibitem{Zhang:2009ut}
L.~Zhang, C.~Weniger, L.~Maccione, J.~Redondo and G.~Sigl,
%``Constraining Decaying Dark Matter with Fermi LAT Gamma-rays,''
JCAP \textbf{06}, 027 (2010)
doi:10.1088/1475-7516/2010/06/027
[arXiv:0912.4504 [astro-ph.HE]].
%43 citations counted in INSPIRE as of 01 Dec 2020

%\cite{Cirelli:2009dv}
\bibitem{Cirelli:2009dv}
M.~Cirelli, P.~Panci and P.~D.~Serpico,
%``Diffuse gamma ray constraints on annihilating or decaying Dark Matter after Fermi,''
Nucl. Phys. B \textbf{840}, 284-303 (2010)
doi:10.1016/j.nuclphysb.2010.07.010
[arXiv:0912.0663 [astro-ph.CO]].
%212 citations counted in INSPIRE as of 01 Dec 2020

%\cite{Bell:2010fk}
\bibitem{Bell:2010fk}
N.~F.~Bell, A.~J.~Galea and K.~Petraki,
%``Lifetime Constraints for Late Dark Matter Decay,''
Phys. Rev. D \textbf{82}, 023514 (2010)
doi:10.1103/PhysRevD.82.023514
[arXiv:1004.1008 [astro-ph.HE]].
%33 citations counted in INSPIRE as of 01 Dec 2020

%\cite{Dugger:2010ys}
\bibitem{Dugger:2010ys}
L.~Dugger, T.~E.~Jeltema and S.~Profumo,
%``Constraints on Decaying Dark Matter from Fermi Observations of Nearby Galaxies and Clusters,''
JCAP \textbf{12}, 015 (2010)
doi:10.1088/1475-7516/2010/12/015
[arXiv:1009.5988 [astro-ph.HE]].
%118 citations counted in INSPIRE as of 01 Dec 2020

%\cite{Cirelli:2012ut}
\bibitem{Cirelli:2012ut}
M.~Cirelli, E.~Moulin, P.~Panci, P.~D.~Serpico and A.~Viana,
%``Gamma ray constraints on Decaying Dark Matter,''
Phys. Rev. D \textbf{86}, 083506 (2012)
doi:10.1103/PhysRevD.86.083506
[arXiv:1205.5283 [astro-ph.CO]].
%120 citations counted in INSPIRE as of 01 Dec 2020

%\cite{Murase:2012xs}
\bibitem{Murase:2012xs}
K.~Murase and J.~F.~Beacom,
%``Constraining Very Heavy Dark Matter Using Diffuse Backgrounds of Neutrinos and Cascaded Gamma Rays,''
JCAP \textbf{10}, 043 (2012)
doi:10.1088/1475-7516/2012/10/043
[arXiv:1206.2595 [hep-ph]].
%105 citations counted in INSPIRE as of 27 Nov 2020

%\cite{Mambrini:2015sia}
\bibitem{Mambrini:2015sia}
Y.~Mambrini, S.~Profumo and F.~S.~Queiroz,
%``Dark Matter and Global Symmetries,''
Phys. Lett. B \textbf{760}, 807-815 (2016)
doi:10.1016/j.physletb.2016.07.076
[arXiv:1508.06635 [hep-ph]].
%95 citations counted in INSPIRE as of 02 Dec 2020

%\cite{Essig:2013goa}
\bibitem{Essig:2013goa}
R.~Essig, E.~Kuflik, S.~D.~McDermott, T.~Volansky and K.~M.~Zurek,
%``Constraining Light Dark Matter with Diffuse X-Ray and Gamma-Ray Observations,''
JHEP \textbf{11}, 193 (2013)
doi:10.1007/JHEP11(2013)193
[arXiv:1309.4091 [hep-ph]].
%183 citations counted in INSPIRE as of 10 Dec 2020

%\cite{Gruber:1999yr}
\bibitem{Gruber:1999yr}
D.~E.~Gruber, J.~L.~Matteson, L.~E.~Peterson and G.~V.~Jung,
%``The spectrum of diffuse cosmic hard x-rays measured with heao-1,''
Astrophys. J. \textbf{520}, 124 (1999)
doi:10.1086/307450
[arXiv:astro-ph/9903492 [astro-ph]].
%254 citations counted in INSPIRE as of 10 Dec 2020

%\cite{Bouchet:2008rp}
\bibitem{Bouchet:2008rp}
L.~Bouchet, E.~Jourdain, J.~P.~Roques, A.~Strong, R.~Diehl, F.~Lebrun and R.~Terrier,
%``INTEGRAL SPI All-Sky View in Soft Gamma Rays: Study of Point Source and Galactic Diffuse Emissions,''
Astrophys. J. \textbf{679}, 1315 (2008)
doi:10.1086/529489
[arXiv:0801.2086 [astro-ph]].
%76 citations counted in INSPIRE as of 01 Dec 2020

%\cite{Weidenspointner:2000aq}
\bibitem{Weidenspointner:2000aq}
G.~Weidenspointner, M.~Varendorff, U.~Oberlack, D.~Morris, S.~Plueschke, R.~Diehl, S.~C.~Kappadath, M.~McConnell, J.~Ryan and V.~Schoenfelder, \textit{et al.}
%``The comptel instrumental line background,''
AIP Conf. Proc. \textbf{510}, no.1, 581-585 (2000)
doi:10.1063/1.1303269
[arXiv:astro-ph/0012332 [astro-ph]].
%19 citations counted in INSPIRE as of 27 Nov 2020

%\cite{Strong:2004de}
\bibitem{Strong:2004de}
A.~W.~Strong, I.~V.~Moskalenko and O.~Reimer,
%``Diffuse galactic continuum gamma rays. A Model compatible with EGRET data and cosmic-ray measurements,''
Astrophys. J. \textbf{613}, 962-976 (2004)
doi:10.1086/423193
[arXiv:astro-ph/0406254 [astro-ph]].
%487 citations counted in INSPIRE as of 10 Dec 2020

%\cite{Slatyer:2016qyl}
\bibitem{Slatyer:2016qyl}
T.~R.~Slatyer and C.~L.~Wu,
%``General Constraints on Dark Matter Decay from the Cosmic Microwave Background,''
Phys. Rev. D \textbf{95}, no.2, 023010 (2017)
doi:10.1103/PhysRevD.95.023010
[arXiv:1610.06933 [astro-ph.CO]].
%121 citations counted in INSPIRE as of 11 Dec 2020


%\cite{Holdom:1985ag}
\bibitem{Holdom:1985ag}
B.~Holdom,
%``Two U(1)'s and Epsilon Charge Shifts,''
Phys. Lett. B \textbf{166}, 196-198 (1986)
doi:10.1016/0370-2693(86)91377-8;
  %%CITATION = doi:10.1016/0370-2693(86)91377-8;%%
  %1599 citations counted in INSPIRE as of 23 Feb 2020;
%\bibitem{Holdom:1991}
 % B. Holdom,
  Phys.\ Lett.\ B {\bf 259}, 329 (1991).
  doi:10.1016/0370-2693(91)90836-F
  %%CITATION = doi:10.1016/0370-2693(91)90836-F;%%
  %165 citations counted in INSPIRE as of 18 May 2018
%

%\cite{Kors:2004dx}
\bibitem{Kors:2004dx}
B.~Kors and P.~Nath,
%``A Stueckelberg extension of the standard model,''
Phys. Lett. B \textbf{586}, 366-372 (2004)
doi:10.1016/j.physletb.2004.02.051
[arXiv:hep-ph/0402047 [hep-ph]];
%196 citations counted in INSPIRE as of 15 Oct 2020
JHEP \textbf{07}, 069 (2005)
doi:10.1088/1126-6708/2005/07/069.
%[arXiv:hep-ph/0503208 [hep-ph]].
%149 citations counted in INSPIRE as of 11 Nov 2020


%\cite{Cheung:2007ut}
\bibitem{Cheung:2007ut}
K.~Cheung and T.~C.~Yuan,
%``Hidden fermion as milli-charged dark matter in Stueckelberg Z- prime model,''
JHEP \textbf{03}, 120 (2007)
doi:10.1088/1126-6708/2007/03/120
[arXiv:hep-ph/0701107 [hep-ph]];
%91 citations counted in INSPIRE as of 25 Oct 2020
%\cite{Feldman:2008xs}
%\bibitem{Feldman:2008xs}
D.~Feldman, Z.~Liu and P.~Nath,
%``PAMELA Positron Excess as a Signal from the Hidden Sector,''
Phys. Rev. D \textbf{79}, 063509 (2009)
doi:10.1103/PhysRevD.79.063509
[arXiv:0810.5762 [hep-ph]].
%218 citations counted in INSPIRE as of 11 Dec 2020


%\cite{Feldman:2007wj}
\bibitem{Feldman:2007wj}
D.~Feldman, Z.~Liu and P.~Nath,
%``The Stueckelberg Z-prime Extension with Kinetic Mixing and Milli-Charged Dark Matter From the Hidden Sector,''
Phys. Rev. D \textbf{75}, 115001 (2007)
doi:10.1103/PhysRevD.75.115001
[arXiv:hep-ph/0702123 [hep-ph]].
%292 citations counted in INSPIRE as of 11 Dec 2020

\bibitem{mSUGRA}
A.~H.~Chamseddine, R.~Arnowitt and P.~Nath,
  %``Locally Supersymmetric Grand Unification,''
  Phys.\ Rev.\ Lett.\  {\bf 49} (1982) 970;
  %%CITATION = PRLTA,49,970;%%
  P.~Nath, R.~L.~Arnowitt and A.~H.~Chamseddine,
  %``Gauge Hierarchy In Supergravity Guts,''
  Nucl.\ Phys.\  B {\bf 227}, 121 (1983);
  %%CITATION = NUPHA,B227,121;%%
 L.~J.~Hall, J.~D.~Lykken and S.~Weinberg,
  %``Supergravity as the Messenger of Supersymmetry Breaking,''
  Phys.\ Rev.\ D {\bf 27}, 2359 (1983).
  doi:10.1103/PhysRevD.27.2359
  %%CITATION = doi:10.1103/PhysRevD.27.2359;%%
  %1413 citations counted in INSPIRE as of 07 Apr 2017

\bibitem{nonuni-gaugino}
 A.~Corsetti and P.~Nath,
  %``Gaugino mass nonuniversality and dark matter in SUGRA, strings and D  brane
  %models,''
  Phys.\ Rev.\  D {\bf 64}, 125010 (2001);
 % [arXiv:hep-ph/0003186].
  %%CITATION = PHRVA,D64,125010;%%
U.~Chattopadhyay and P.~Nath,
  %``b - tau unification, g(mu)-2, the b --> s + gamma constraint and
  %nonuniversalities,''
  Phys.\ Rev.\  D {\bf 65}, 075009 (2002);
%  [arXiv:hep-ph/0110341].
  %%CITATION = PHRVA,D65,075009;%%
  A.~Birkedal-Hansen and B.~D.~Nelson,
  %``Relic neutralino densities and detection rates with nonuniversal gaugino
  %masses,''
  Phys.\ Rev.\  D {\bf 67}, 095006 (2003);
 % [arXiv:hep-ph/0211071].
  %%CITATION = PHRVA,D67,095006;%%
  H.~Baer, A.~Mustafayev, E.~K.~Park, S.~Profumo and X.~Tata,
  %``Mixed higgsino dark matter from a reduced SU(3) gaugino mass:  Consequences
  %for dark matter and collider searches,''
  JHEP {\bf 0604}, 041 (2006);
  K.~Choi and H.~P.~Nilles
  %``The gaugino code,''
  JHEP {\bf 0704} (2007) 006;
  I.~Gogoladze, R.~Khalid, N.~Okada and Q.~Shafi,
  %``Soft Probes of SU(5) Unification,''
  arXiv:0811.1187 [hep-ph];
  %%CITATION = ARXIV:0811.1187;%
  S.~P.~Martin,
  %``Non-universal gaugino masses from non-singlet F-terms in non-minimal unified models,''
  Phys.\ Rev.\ D {\bf 79}, 095019 (2009)
  doi:10.1103/PhysRevD.79.095019
  [arXiv:0903.3568 [hep-ph]].
  %%CITATION = doi:10.1103/PhysRevD.79.095019;%%
  %108 citations counted in INSPIRE as of 24 Oct 2019

%\cite{Herrmann:2009mp}
\bibitem{Herrmann:2009mp}
B.~Herrmann, M.~Klasen and K.~Kovarik,
%``SUSY-QCD effects on neutralino dark matter annihilation beyond scalar or gaugino mass unification,''
Phys. Rev. D \textbf{80}, 085025 (2009)
doi:10.1103/PhysRevD.80.085025
[arXiv:0907.0030 [hep-ph]].
%46 citations counted in INSPIRE as of 27 Nov 2020

%\cite{Ibrahim:2004gb}
\bibitem{Ibrahim:2004gb}
T.~Ibrahim and P.~Nath,
%``Effective Lagrangian for anti-q \textasciitilde{}q-prime(i) Chi+(j), anti-q \textasciitilde{}q-prime(i) Chi0(j) interactions and fermionic decays of the squarks with CP phases,''
Phys. Rev. D \textbf{71}, 055007 (2005)
doi:10.1103/PhysRevD.71.055007
[arXiv:hep-ph/0411272 [hep-ph]].
%10 citations counted in INSPIRE as of 27 Nov 2020

%\cite{Navarro:1995iw}
\bibitem{Navarro:1995iw}
J.~F.~Navarro, C.~S.~Frenk and S.~D.~M.~White,
%``The Structure of cold dark matter halos,''
Astrophys. J. \textbf{462}, 563-575 (1996)
doi:10.1086/177173
[arXiv:astro-ph/9508025 [astro-ph]].
%5044 citations counted in INSPIRE as of 11 Dec 2020

%\cite{Sjostrand:2007gs}
\bibitem{Sjostrand:2007gs}
T.~Sjostrand, S.~Mrenna and P.~Z.~Skands,
%``A Brief Introduction to PYTHIA 8.1,''
Comput. Phys. Commun. \textbf{178}, 852-867 (2008)
doi:10.1016/j.cpc.2008.01.036
[arXiv:0710.3820 [hep-ph]].
%5708 citations counted in INSPIRE as of 11 Dec 2020

%\cite{Amoroso:2018qga}
\bibitem{Amoroso:2018qga}
S.~Amoroso, S.~Caron, A.~Jueid, R.~Ruiz de Austri and P.~Skands,
%``Estimating QCD uncertainties in Monte Carlo event generators for gamma-ray dark matter searches,''
JCAP \textbf{05}, 007 (2019)
doi:10.1088/1475-7516/2019/05/007
[arXiv:1812.07424 [hep-ph]].
%12 citations counted in INSPIRE as of 25 Jan 2021


%\cite{Ciafaloni:2010ti}
\bibitem{Ciafaloni:2010ti}
P.~Ciafaloni, D.~Comelli, A.~Riotto, F.~Sala, A.~Strumia and A.~Urbano,
%``Weak Corrections are Relevant for Dark Matter Indirect Detection,''
JCAP \textbf{03}, 019 (2011)
doi:10.1088/1475-7516/2011/03/019
[arXiv:1009.0224 [hep-ph]].
%298 citations counted in INSPIRE as of 11 Dec 2020

%\cite{Cirelli:2010xx}
\bibitem{Cirelli:2010xx}
M.~Cirelli, G.~Corcella, A.~Hektor, G.~Hutsi, M.~Kadastik, P.~Panci, M.~Raidal, F.~Sala and A.~Strumia,
%``PPPC 4 DM ID: A Poor Particle Physicist Cookbook for Dark Matter Indirect Detection,''
JCAP \textbf{03}, 051 (2011)
[erratum: JCAP \textbf{10}, E01 (2012)]
doi:10.1088/1475-7516/2012/10/E01
[arXiv:1012.4515 [hep-ph]].
%774 citations counted in INSPIRE as of 11 Dec 2020

%\cite{Buch:2015iya}
\bibitem{Buch:2015iya}
J.~Buch, M.~Cirelli, G.~Giesen and M.~Taoso,
%``PPPC 4 DM secondary: A Poor Particle Physicist Cookbook for secondary radiation from Dark Matter,''
JCAP \textbf{09}, 037 (2015)
doi:10.1088/1475-7516/2015/9/037
[arXiv:1505.01049 [hep-ph]].
%39 citations counted in INSPIRE as of 27 Nov 2020

%\cite{Ginzburg:1990sk}
\bibitem{Ginzburg:1990sk}
V.~S.~Berezinsky, S.~V.~Bulanov, V.~A.~Dogiel, V.~L.~Ginzburg and V.~S.~Ptuskin,
%``Astrophysics of cosmic rays,''
%141 citations counted in INSPIRE as of 03 Dec 2020

%\cite{Ibarra:2009bm}
\bibitem{Ibarra:2009bm}
A.~Ibarra, A.~Ringwald, D.~Tran and C.~Weniger,
%``Cosmic Rays from Leptophilic Dark Matter Decay via Kinetic Mixing,''
JCAP \textbf{08}, 017 (2009)
doi:10.1088/1475-7516/2009/08/017
[arXiv:0903.3625 [hep-ph]].
%87 citations counted in INSPIRE as of 27 Nov 2020

%\cite{Boudaud:2014qra}
\bibitem{Boudaud:2014qra}
M.~Boudaud, M.~Cirelli, G.~Giesen and P.~Salati,
%``A fussy revisitation of antiprotons as a tool for Dark Matter searches,''
JCAP \textbf{05}, 013 (2015)
doi:10.1088/1475-7516/2015/05/013
[arXiv:1412.5696 [astro-ph.HE]].
%50 citations counted in INSPIRE as of 10 Dec 2020

%\cite{Abdo:2010nc}
\bibitem{Abdo:2010nc}
A.~A.~Abdo, M.~Ackermann, M.~Ajello, W.~B.~Atwood, L.~Baldini, J.~Ballet, G.~Barbiellini, D.~Bastieri, K.~Bechtol and R.~Bellazzini, \textit{et al.}
%``Fermi LAT Search for Photon Lines from 30 to 200 GeV and Dark Matter Implications,''
Phys. Rev. Lett. \textbf{104}, 091302 (2010)
doi:10.1103/PhysRevLett.104.091302
[arXiv:1001.4836 [astro-ph.HE]].
%227 citations counted in INSPIRE as of 01 Dec 2020

%\cite{Ackermann:2015lka}
\bibitem{Ackermann:2015lka}
M.~Ackermann \textit{et al.} [Fermi-LAT],
%``Updated search for spectral lines from Galactic dark matter interactions with pass 8 data from the Fermi Large Area Telescope,''
Phys. Rev. D \textbf{91}, no.12, 122002 (2015)
doi:10.1103/PhysRevD.91.122002
[arXiv:1506.00013 [astro-ph.HE]].
%273 citations counted in INSPIRE as of 11 Dec 2020

%\cite{Ackermann:2014usa}
\bibitem{Ackermann:2014usa}
M.~Ackermann \textit{et al.} [Fermi-LAT],
%``The spectrum of isotropic diffuse gamma-ray emission between 100 MeV and 820 GeV,''
Astrophys. J. \textbf{799}, 86 (2015)
doi:10.1088/0004-637X/799/1/86
[arXiv:1410.3696 [astro-ph.HE]].
%533 citations counted in INSPIRE as of 11 Dec 2020

%\cite{Blanco:2018esa}
\bibitem{Blanco:2018esa}
C.~Blanco and D.~Hooper,
%``Constraints on Decaying Dark Matter from the Isotropic Gamma-Ray Background,''
JCAP \textbf{03}, 019 (2019)
doi:10.1088/1475-7516/2019/03/019
[arXiv:1811.05988 [astro-ph.HE]].
%29 citations counted in INSPIRE as of 11 Dec 2020

%\cite{Porod:2003um}
\bibitem{Porod:2003um}
W.~Porod,
%``SPheno, a program for calculating supersymmetric spectra, SUSY particle decays and SUSY particle production at e+ e- colliders,''
Comput. Phys. Commun. \textbf{153}, 275-315 (2003)
doi:10.1016/S0010-4655(03)00222-4
[arXiv:hep-ph/0301101 [hep-ph]].
%995 citations counted in INSPIRE as of 10 Dec 2020

%\cite{Porod:2011nf}
\bibitem{Porod:2011nf}
W.~Porod and F.~Staub,
%``SPheno 3.1: Extensions including flavour, CP-phases and models beyond the MSSM,''
Comput. Phys. Commun. \textbf{183}, 2458-2469 (2012)
doi:10.1016/j.cpc.2012.05.021
[arXiv:1104.1573 [hep-ph]].
%633 citations counted in INSPIRE as of 10 Dec 2020


%\cite{Belanger:2014vza}
\bibitem{Belanger:2014vza}
G.~B\'elanger, F.~Boudjema, A.~Pukhov and A.~Semenov,
%``micrOMEGAs4.1: two dark matter candidates,''
Comput. Phys. Commun. \textbf{192}, 322-329 (2015)
doi:10.1016/j.cpc.2015.03.003
[arXiv:1407.6129 [hep-ph]].
%348 citations counted in INSPIRE as of 01 Dec 2020

%\cite{Aghanim:2018eyx}
\bibitem{Aghanim:2018eyx}
N.~Aghanim \textit{et al.} [Planck],
%``Planck 2018 results. VI. Cosmological parameters,''
Astron. Astrophys. \textbf{641}, A6 (2020)
doi:10.1051/0004-6361/201833910
[arXiv:1807.06209 [astro-ph.CO]].
%4062 citations counted in INSPIRE as of 12 Dec 2020

%\cite{Goodsell:2009xc}
\bibitem{Goodsell:2009xc}
M.~Goodsell, J.~Jaeckel, J.~Redondo and A.~Ringwald,
%``Naturally Light Hidden Photons in LARGE Volume String Compactifications,''
JHEP \textbf{11}, 027 (2009)
doi:10.1088/1126-6708/2009/11/027
[arXiv:0909.0515 [hep-ph]].
%273 citations counted in INSPIRE as of 22 Jan 2021

%\cite{Hall:2009bx}
\bibitem{Hall:2009bx}
L.~J.~Hall, K.~Jedamzik, J.~March-Russell and S.~M.~West,
%``Freeze-In Production of FIMP Dark Matter,''
JHEP \textbf{03}, 080 (2010)
doi:10.1007/JHEP03(2010)080
[arXiv:0911.1120 [hep-ph]].
%652 citations counted in INSPIRE as of 11 Dec 2020

%\cite{Aboubrahim:2019kpb}
\bibitem{Aboubrahim:2019kpb}
A.~Aboubrahim, W.~Z.~Feng and P.~Nath,
%``A long-lived stop with freeze-in and freeze-out dark matter in the hidden sector,''
JHEP \textbf{02}, 118 (2020)
doi:10.1007/JHEP02(2020)118
[arXiv:1910.14092 [hep-ph]].
%4 citations counted in INSPIRE as of 27 Nov 2020

%\cite{Aboubrahim:2020wah}
\bibitem{Aboubrahim:2020wah}
A.~Aboubrahim, W.~Z.~Feng and P.~Nath,
%``Expanding the parameter space of natural supersymmetry,''
JHEP \textbf{04}, 144 (2020)
doi:10.1007/JHEP04(2020)144
[arXiv:2003.02267 [hep-ph]].
%3 citations counted in INSPIRE as of 27 Nov 2020

%\cite{Aboubrahim:2020lnr}
\bibitem{Aboubrahim:2020lnr}
A.~Aboubrahim, W.~Z.~Feng, P.~Nath and Z.~Y.~Wang,
%``Self-interacting hidden sector dark matter and small scale galaxy structure anomalies,''
[arXiv:2008.00529 [hep-ph]].
%1 citations counted in INSPIRE as of 14 Dec 2020


%\cite{Baer:2006rs}
\bibitem{Baer:2006rs}
H.~Baer and X.~Tata, \textit{Weak scale supersymmetry: From superfields to scattering events}. Cambridge University Press, 5, 2006.
%133 citations counted in INSPIRE as of 22 Jan 2021


%\cite{Aharonian:2003xh}
\bibitem{Aharonian:2003xh}
F.~A.~Aharonian \textit{et al.} [HEGRA],
%``Search for TeV gamma-ray emission from the Andromeda galaxy,''
Astron. Astrophys. \textbf{400}, 153-159 (2003)
doi:10.1051/0004-6361:20021895
[arXiv:astro-ph/0302347 [astro-ph]].
%22 citations counted in INSPIRE as of 01 Dec 2020


%\cite{Aprile:2020tmw}
\bibitem{Aprile:2020tmw}
E.~Aprile \textit{et al.} [XENON],
%``Excess electronic recoil events in XENON1T,''
Phys. Rev. D \textbf{102}, no.7, 072004 (2020)
doi:10.1103/PhysRevD.102.072004
[arXiv:2006.09721 [hep-ex]].
%166 citations counted in INSPIRE as of 07 Dec 2020


%\cite{Aboubrahim:2020iwb}
\bibitem{Aboubrahim:2020iwb}
A.~Aboubrahim, M.~Klasen and P.~Nath,
%``Xenon-1T excess as a possible signal of a sub-GeV hidden sector dark matter,''
[arXiv:2011.08053 [hep-ph]].
%1 citations counted in INSPIRE as of 14 Dec 2020

%\cite{Sierra:2009zq}
\bibitem{Sierra:2009zq}
D.~Aristizabal Sierra, D.~Restrepo and O.~Zapata,
%``Decaying Neutralino Dark Matter in Anomalous U(1)(H) Models,''
Phys. Rev. D \textbf{80}, 055010 (2009)
doi:10.1103/PhysRevD.80.055010
[arXiv:0907.0682 [hep-ph]].
%24 citations counted in INSPIRE as of 24 Jan 2021

%\cite{Aboubrahim:2019qpc}
\bibitem{Aboubrahim:2019qpc}
A.~Aboubrahim and P.~Nath,
%``Detecting hidden sector dark matter at HL-LHC and HE-LHC via long-lived stau decays,''
Phys. Rev. D \textbf{99}, no.5, 055037 (2019)
doi:10.1103/PhysRevD.99.055037
[arXiv:1902.05538 [hep-ph]].
%11 citations counted in INSPIRE as of 17 Feb 2021

%\cite{Aprile:2018dbl}
\bibitem{Aprile:2018dbl}
E.~Aprile \textit{et al.} [XENON],
%``Dark Matter Search Results from a One Ton-Year Exposure of XENON1T,''
Phys. Rev. Lett. \textbf{121}, no.11, 111302 (2018)
doi:10.1103/PhysRevLett.121.111302
[arXiv:1805.12562 [astro-ph.CO]].
%1043 citations counted in INSPIRE as of 17 Feb 2021

%\cite{Aaboud:2018sua}
\bibitem{Aaboud:2018sua}
M.~Aaboud \textit{et al.} [ATLAS],
%``Search for chargino-neutralino production using recursive jigsaw reconstruction in final states with two or three charged leptons in proton-proton collisions at $\sqrt{s}=13$ TeV with the ATLAS detector,''
Phys. Rev. D \textbf{98}, no.9, 092012 (2018)
doi:10.1103/PhysRevD.98.092012
[arXiv:1806.02293 [hep-ex]].
%67 citations counted in INSPIRE as of 17 Feb 2021

%\cite{Aad:2019qnd}
\bibitem{Aad:2019qnd}
G.~Aad \textit{et al.} [ATLAS],
%``Searches for electroweak production of supersymmetric particles with compressed mass spectra in $\sqrt{s}=$ 13 TeV $pp$ collisions with the ATLAS detector,''
Phys. Rev. D \textbf{101}, no.5, 052005 (2020)
doi:10.1103/PhysRevD.101.052005
[arXiv:1911.12606 [hep-ex]].
%52 citations counted in INSPIRE as of 17 Feb 2021

%\cite{ATLAS:2020ckz}
\bibitem{ATLAS:2020ckz}
 [ATLAS],
%``Search for chargino-neutralino pair production in final states with three leptons and missing transverse momentum in $\sqrt{s}$ = 13 TeV p-p collisions with the ATLAS detector,''
ATLAS-CONF-2020-015.
%6 citations counted in INSPIRE as of 17 Feb 2021

%\cite{Aad:2015eda}
\bibitem{Aad:2015eda}
G.~Aad \textit{et al.} [ATLAS],
%``Search for the electroweak production of supersymmetric particles in $\sqrt{s}$=8 TeV $pp$ collisions with the ATLAS detector,''
Phys. Rev. D \textbf{93}, no.5, 052002 (2016)
doi:10.1103/PhysRevD.93.052002
[arXiv:1509.07152 [hep-ex]].
%136 citations counted in INSPIRE as of 17 Feb 2021

%\cite{Aaboud:2018jiw}
\bibitem{Aaboud:2018jiw}
M.~Aaboud \textit{et al.} [ATLAS],
%``Search for electroweak production of supersymmetric particles in final states with two or three leptons at $\sqrt{s}=13\,$TeV with the ATLAS detector,''
Eur. Phys. J. C \textbf{78}, no.12, 995 (2018)
doi:10.1140/epjc/s10052-018-6423-7
[arXiv:1803.02762 [hep-ex]].
%147 citations counted in INSPIRE as of 17 Feb 2021



\end{thebibliography}
\end{document}